\documentclass[12pt]{article}

\usepackage{amsmath,amsfonts}
\usepackage{amssymb}
\usepackage{epsfig,psfrag,cite}
\usepackage[usenames,dvips]{color}
\usepackage{comment}
\usepackage{multirow,axodraw}

\usepackage{fullpage}

\makeatletter
\@addtoreset{equation}{section}
\makeatother

\newcommand{\be}{\begin{equation}}
\newcommand{\ee}{\end{equation}}
\newcommand{\bea}{\begin{eqnarray}}
\newcommand{\eea}{\end{eqnarray}}

\newcommand{\tr}{\operatorname{tr}}
\newcommand{\diag}{\operatorname{diag}}

\newcommand{\nn}{\nonumber}

\newcommand{\M}{\mathcal M}

\begin{document}

\thispagestyle{empty}

\begin{center}
\hfill CPHT-RR064.0911\\
\hfill UAB-FT-696

\begin{center}

\vspace{.5cm}

{\LARGE\bf  Flavor Phenomenology 
in General \\ 5D Warped Spaces}

\end{center}

\vspace{1.cm}

{\bf Joan A. Cabrer$^{\,a}$, Gero von Gersdorff$^{\;b}$ 
and Mariano Quir\'os$^{\,c,\,d}$}\\

\vspace{1.cm}
${}^a\!\!$ {\em {Institut de F\'isica d'Altes Energies  UAB,
08193 Bellaterra, Barcelona, Spain}}

\vspace{.1cm}

${}^b\!\!$ {\em {Centre de Physique Th\'eorique, \'Ecole Polytechnique and CNRS
F-91128 Palaiseau, France}}

\vspace{.1cm}

${}^c\!\!$ {\em {Department of Physics, University of Notre Dame,
Notre Dame, IN 46556, USA}}

\vspace{.1cm}

${}^d\!\!$ {\em {Instituci\'o Catalana de Recerca i Estudis  
Avan\c{c}ats (ICREA) and\\ Institut de F\'isica d'Altes Energies UAB 
08193 Bellaterra, Barcelona, Spain}}

\end{center}

\vspace{0.8cm}

\centerline{\bf Abstract}
\vspace{2 mm}
\begin{quote}\small
We have considered a general 5D warped model with SM fields propagating in the bulk and computed explicit expressions for oblique and non-oblique electroweak observables as well as for flavor and $CP$ violating effective four-fermion operators. We have compared the resulting lower bounds on the Kaluza Klein (KK) scale in the RS model and a recently proposed model with a metric modified towards the IR brane, which is consistent with oblique parameters without the need for a custodial symmetry. We have randomly generated 40,000 sets of $\mathcal O(1)$ 5D Yukawa couplings and made a fit of the quark masses and CKM matrix elements in both models. This method allows to identify the percentage of points consistent with a given KK mass, which in turn provides us with a measure for the required fine-tuning. Comparison with current experimental data on $R_b$, FCNC and $CP$ violating operators exhibits an improved behavior of our model with respect to the RS model. In particular, allowing  10\% fine-tuning the combined results point towards upper bounds on the KK gauge boson masses around 3.3 TeV in our model as compared with 13 TeV in the RS model. One reason for this improvement is that fermions in our model are shifted, with respect to fermions in the RS model, towards the UV brane thus decreasing the strength of the modifications of electroweak observables.
\end{quote}

\vfill

 \newpage

\section{Introduction}

The Standard Model (SM) of electroweak (EW) interactions suffers from a naturalness problem, as the mass of the Higgs and  its vacuum expectation value (VEV) are sensitive to the ultraviolet cutoff. This is known as the hierarchy problem and a number of SM extensions have been proposed with the aim of solving it. One of the most interesting solutions was originally proposed by Randall and Sundrum (RS)~\cite{Randall:1999ee} and it is based on a five-dimensional (5D) space-time with Anti de Sitter (AdS) metric,
\be
ds^2=e^{-2A(y)}\eta_{\mu\nu}dx^\mu dx^\nu+dy^2,\quad A(y)=ky\,,
\label{RSmetric}
\ee
where $k\sim M_{P\ell}$ is the AdS inverse curvature. The model has two boundaries, the ultraviolet (UV) brane, located at $y=0$, and the infrared (IR) brane, located at $y=y_1$. The brane distance can be stabilized by an extra bulk Goldberger-Wise (GW) scalar field $\phi$~\cite{Goldberger:1999uk}. The original model had all the SM (and in particular the Higgs field) contained in the IR boundary such that the Planck Higgs mass is redshifted to the TeV scale by the warp factor and the hierarchy problem is solved provided that $ky_1\simeq 35$.

However the SM should not necessarily be localized in the IR boundary. In fact if the fermions (and gauge bosons) propagate in the bulk
with $\mathcal O(k)$ 5D Dirac masses the model could furnish a theory of flavor~\cite{Grossman:1999ra}. Moreover if the Higgs propagates in the bulk it can solve the hierarchy problem if it is sufficiently localized towards the IR boundary. In particular assuming a 5D bulk Higgs mass $M^2=a(4-a)k^2$ the solution to the gravitational equations of motion (EOM) of the Higgs field in the AdS background provides a Higgs profile as
\be
h(y)=h(0)e^{aky}\,,
\label{Higgsprofile}
\ee
where the size of $a$ measures the degree of localization of the Higgs towards the IR boundary.  Given the holographic interpretation of the parameter $a$ as the dimension of the Higgs condensate, $a=\dim(\mathcal O_H)$,
 the solution to the hierarchy problem requires  the lower bound $a>2$. 

However, confronting the electroweak precision observables (EWPO) in the RS model with experimental data, or electroweak precision tests (EWPT), translates into lower bounds on the Kaluza-Klein (KK) gauge bosons mass, $m_{KK}$, which are outside the LHC reach and thus recreate a little hierarchy problem. For instance for an IR localized Higgs with a mass $m_H=115$ GeV,  $m_{KK}\gtrsim 13$ TeV, while for a bulk Higgs with $a\gtrsim 2$ one obtains $m_{KK}\gtrsim 7.5$ TeV \cite{Cabrer:2011vu}. The origin of this large bounds can be traced back to the scaling dependence of the $T$ (i.e.~$\Delta\rho$) parameter with the compactification volume, in particular $\alpha T\sim ky_1$. In order to fix the problem different solutions have been proposed in the literature.
\begin{itemize}
\item
One possible solution is to introduce large IR localized gauge boson kinetic terms~\cite{Carena:2002dz}. However such large kinetic terms cannot come from radiative corrections and their origin should be traced back to unknown UV physics.
\item
Another very interesting possibility is to enlarge the SM gauge group and particle content by introducing an extra gauge custodial symmetry in the bulk protecting the $T$ parameter at the tree level~\cite{Agashe:2003zs}. In custodial symmetry models there are extra light modes on top of the SM ones.
\item
A third possibility is to generalize the AdS metric in the IR with a strong deformation of conformality such that the coupling of EW KK modes to the Higgs is suppressed~\cite{Cabrer:2010si}, consequently softening the bounds coming from oblique EWPO. This solution allows for a minimal 5D extension of the SM.
\end{itemize}

The last possibility was analyzed by the present authors and confronted with oblique EWPO's in Refs.~\cite{Cabrer:2010si,Cabrer:2011vu} yielding bounds on $m_{KK}$ as low as $\sim 1$ TeV for $m_H\simeq 115$ GeV. We used a GW stabilizing field $\phi$ providing a metric~\cite{Cabrer:2009we}
\be
A(y)=k y-\frac{1}{\nu^2}\log\left(1-\frac{y}{y_1+\Delta}\right)\,,
\label{ourmetric}
\ee
with a spurious singularity located at $y_s=y_1+\Delta$, outside the physical interval. The dynamics of the stabilizing field $\phi$ fixes the values of $y_1$ (i.e. the warp factor $A(y_1)$) and $\Delta$ by a GW mechanism where the back reaction on the metric is included and generates the metric in Eq.~(\ref{ourmetric}). A bulk 5D Higgs mass is introduced as $M^2=a(a-4-e^{\nu\phi})k^2$ and the solution to the EOM of the Higgs background is given by Eq.~(\ref{Higgsprofile}) while the condition for solving the hierarchy problem reads now as
\be
a\gtrsim a_0=2\frac{A(y_1)}{ky_1}\,.
\label{hierarchy}
\ee

In this paper we will pursue the phenomenological analysis of warped models with bulk fermions describing the flavor in the quark sector by means of different localization (or 5D Dirac masses) for different fermions and no special structure for the 5D Yukawa couplings~\footnote{The absence of a 5D Yukawa hierarchy is sometimes referred to as anarchy.}. It is well known that the different fermion localization generates non-oblique observables, mainly it modifies the $Z\bar bb$ coupling, as well as flavor changing neutral current (FCNC) and CP violating dimension-six operators. We will get bounds on $m_{KK}$ from both $Z\bar bb$ coupling and the flavor violating operators for both models based on the RS metric, Eq.~(\ref{RSmetric}), and on the modified background metric of Eq.~(\ref{ourmetric}). In all cases we will find an improvement on the bounds in the modified background model with respect to the RS model because, in order to fit the quark masses and CKM matrix elements, the fermions in models with metric (\ref{ourmetric}) are shifted towards the UV brane with respect to fermions in RS model. This will translate into milder bounds on $m_{KK}$ as we will see.

The plan of the paper goes as follows. In Sec.~\ref{low} we describe the low energy 4D theory obtained after integrating out the KK weak gauge bosons (Sec.~\ref{KKweak}), the KK gluons  (Sec.~\ref{KKgluons}) and the KK fermions  (Sec.~\ref{secKKfermions}). In Sec.~\ref{quarks} we present an approximation of the quark mass eigenstates and mixing angles, assuming a left-handed hierarchy which is more general than other existing approximations in the literature. We also fit the parameters of the 5D theory to the observed quark masses, mixing angles and $CP$-violating CKM phase by using an anarchic structure on the 5D Yukawa couplings. To this end, we have generated a set of 40,000 random complex 5D Yukawa couplings and made a $\chi^2$ fit of the nine 5D Dirac masses for quarks for both the RS model and the model defined by the metric (\ref{ourmetric}) to reproduce the observed masses and mixings. In Sec.~\ref{EWPT} we give explicit expressions of oblique and non-oblique EWPOs for arbitrary metrics. In particular, we extract the bounds on $m_{KK}$ for the RS model and the model defined by the metric (\ref{ourmetric}) for the set of points randomly generated in Sec.~\ref{quarks}. A similar analysis is done in Sec.~\ref{quarkflavorgeneral} for the bounds obtained from FCNC and CP violating dimension-six operators for both models and the randomly generated set of points used in the previous section. Finally, in Sec.~\ref{conclusions} we present our final conclusions and the  combined bounds from both EWPO's and flavor violation. 

Moreover, we include appendices to present a number of technical details, which the reader more interested in the numerical results than in the details of the calculation can skip. In App.~\ref{4f} the subleading four-fermion terms from integration of electroweak KK modes are explicitly presented. In App.~\ref{app:fermions} we present some details of how to formally integrate out KK fermions in general backgrounds. In App.~\ref{RH} we present explicit expressions for the quark mass eigenstates and mixing angles for the particular case where there is a left-handed \textit{and} a right-handed hierarchy using a starting point the more general expressions of Sec.~\ref{quarks} and found agreement with previously published results. In App.~\ref{EWPO} we review the general procedure for linking the various oblique and non-oblique corrections to measurements.

Notice that the issue of non-oblique EWPO's in the model (\ref{ourmetric}) has recently been addressed in Ref.~\cite{Carmona:2011ib}.  Although we employ different  fermion profiles, we find similar bounds from the $Z\bar b b$ coupling, with slightly more optimistic bounds in the fully anarchic case. Moreover our analysis is quite different and complementary to the one done in~\cite{Carmona:2011ib} as we perform a democratic scan over possible 5D Yukawa couplings deducing the bulk masses needed to reproduce the observed masses and mixings. In that way we are able to associate a probability to a certain choice of bulk masses and hence quantify the fine tuning to achieve a given KK scale that leads to agreement with experimental constraints. The improvement of flavor/$CP$ bounds with modified metrics has recently been noted in the context of soft wall models~\cite{Huber2}. Here we show that a similar improvement can be obtained in the hard wall setup, and again we quantify the amount of fine tuning needed to obtain a satisfactory bound on the KK scale.

\section{The low-energy effective theory}
\label{low}

In this section we would like to present general expressions for oblique and non-oblique corrrections, as well as FCNC operators, for arbitrary profiles for the metric, the Higgs and the fermions. We will first integrate out the KK modes of the weak gauge bosons, which will be relevant for EWPO's, in Sec.~\ref{KKweak}. Dimension six operators generated from integration of the KK modes of gluons will be considered in Sec.~\ref{KKgluons} and those obtained from integration of the KK modes of fermions in Sec.~\ref{secKKfermions}.

\subsection{Integrating out the KK weak gauge bosons}
\label{KKweak}

Let us define the currents
\be
J^A_\mu=\{g\, j_\mu^{W\,a},\,g'\,j_\mu^{Y}\}\,,
\ee
where $A=\{a,Y\}$ and $g$ and $g'$ are the 4D gauge couplings corresponding to $SU(2)_L$ and $U(1)_Y$ respectively. The EOM for the EW gauge bosons are then
\be
D^\mu F^A_{\mu\nu}+ J^{A}_{h\, \nu}+\sum_\psi J^{A}_{\psi\, \nu}=0\,,
\label{eom}
\ee
where $J_h$ stands for the Higgs current and $J_\psi$ for fermion currents, with $\psi=Q,L,u,d,e$ the fermions before EWSB, and for now we suppress flavor indices.

The starting point is the effective Lagrangian after integrating out the KK modes
\be
\mathcal L_{\rm eff}=\mathcal L_{\rm SM}+
 \frac{1}{2}\alpha_{hh}\,J_h\cdot J_h+\sum_\psi\alpha_{h\psi}\, J_h\cdot J_\psi
+\frac{1}{2}\sum_{\psi,\psi'} \alpha_{\psi\psi'}\,J_\psi\cdot J_{\psi'}\,,
\label{eff}
\ee
where the coefficients $\alpha_{XX'}$, with $X,X'=\psi,h$, were computed in Ref.~\cite{Cabrer:2010si}
\be
\alpha_{XX'}=y_1\int e^{2A}(\Omega_X-y/y_1)(\Omega_{X'}-y/y_1)\,.
\ee
The functions $\Omega_X$ are defined as
\be
\Omega_X(y)=\int_0^y dy'\,e^{-n_XA(y')}\,[f_X(y')]^2\,,
\label{Omega}
\ee
with $f_X(y)$ the zero mode wave function, and $n_{X}=2\,(3)$ for scalars (fermions). The normalization condition on the wave functions implies $\Omega_X(y_1)=1$.

We will rewrite this Lagrangian as 
\be
\mathcal L_{\rm eff}=\mathcal L_{\rm SM}+\mathcal L_{\rm oblique}+\mathcal L_{\rm non-oblique}\,,
\ee
with
\bea
\mathcal L_{\rm oblique}&=&
\frac{1}{2}\hat\alpha_{hh}\,J_h\cdot J_{h}
+\,\hat\alpha_{hg}\, [D_\mu F^{\mu\nu}\cdot J_{h\,\nu}]
+\frac{1}{2}\hat\alpha_{gg} \, [D_\mu F^{\mu\nu}]^2\label{oblique}\,,\\
\mathcal L_{\rm non-oblique}&=&
\sum_{\psi} \hat\alpha_{h\psi}\,J_h\cdot J_{\psi}
+\frac{1}{2}\sum_{\psi,\psi'} \hat\alpha_{\psi\psi'}\, J_{\psi}\cdot J_{\psi'}\ ,
\label{non-oblique1}
\eea
which is physically equivalent to the Lagrangian Eq.~(\ref{eff}) by use of Eq.~(\ref{eom}) for arbitrary choice of $\hat \alpha_{hg}$ and $\hat \alpha_{gg}$ as long as the following conditions 
\bea
\alpha_{hh}&=&\hat \alpha_{hh}-2\hat \alpha_{hg}+\hat\alpha_{gg}\,,\nn\\
\alpha_{h\psi}&=&\hat \alpha_{h\psi}-\hat\alpha_{hg}+\hat\alpha_{gg}\,,\nn\\
\alpha_{\psi\psi'}&=&\hat\alpha_{\psi\psi'}+\hat\alpha_{gg}\,,
\eea
hold. We can now transform away the non-oblique corrections for near UV localized (mostly elementary) fermions such as first and second generation leptons (which have $\Omega\approx 1$) so all the new physics for them is encoded in the oblique parameters.  
We will refer to this basis as the "oblique basis" and use it from now on. In order to achieve $\hat\alpha_{h\psi}\approx 0$ and $\hat \alpha_{\psi\psi'}\approx 0$ for the elementary fields
a good choice is thus
\bea
\hat\alpha_{hg}&=&y_1\int e^{2A}(1-\Omega_h)(1-y/y_1)\,,\nn\\
\hat\alpha_{gg}&=&y_1\int e^{2A}(1-y/y_1)^2\,,
\eea
which leads to
\begin{eqnarray}
\label{alphahat}
\hat\alpha_{hh}&=&\alpha_{hh}+2\hat\alpha_{hg}-\hat\alpha_{gg}=y_1\int e^{2A}(1-\Omega_h)^2\,,
\nonumber\\
\hat \alpha_{h\psi}&=&\alpha_{h\psi}+\hat \alpha_{hg}-\hat\alpha_{gg}
  =y_1\int e^{2A}(\Omega_h-y/y_1)(\Omega_\psi-1)\,,\\
\hat\alpha_{\psi\psi'}&=&\alpha_{\psi\psi'}-\hat\alpha_{gg}
  =y_1\int e^{2A}[(\Omega_\psi-y/y_1)(\Omega_{\psi'}-y/y_1)-(1-y/y_1)^2]\,.\nonumber
\end{eqnarray}

It is clear from Eq.~(\ref{alphahat}) that, for fermions strictly localized on the UV brane ($\Omega_\psi=1$), $\hat\alpha_{h\psi}$ and $\hat\alpha_{\psi\psi'}$ vanish. Consequently fermions that are only near UV localized will still have strongly suppressed non-oblique corrections. 
The oblique Lagrangian in (\ref{oblique}) gives rise to the $(T,S,W,Y)$ parameters~\cite{Peskin:1991sw,Barbieri:2004qk} while the first term of the non-oblique Lagrangian in (\ref{non-oblique1}) contributes to modified $Z$ and $W$ couplings to fermions as we will describe in detail in Sec.~\ref{EWPT}. The second term of the non-oblique Lagrangian generates flavor violating four-fermion operators as we will describe in App.~\ref{4f}. However the corresponding effects will be subleading with respect to those induced by integration of KK gluons as explained in Sec.~\ref{KKgluons}.

\subsection{Integrating out the KK-gluons}
\label{KKgluons}

Integrating out the KK gluons we obtain
\be
\mathcal L=\frac{g_s^2}{2}\sum_{\psi,\psi'} \alpha_{\psi\psi'}\, 
\bar\psi \gamma^\mu\lambda^a\psi\, \bar\psi'\gamma^\mu\lambda^a\psi'\,,
\ee
where here $\psi$ runs over the quarks ($u_{L,R},d_{L,R}$) and $\lambda^a$ are the $SU(3)$ matrices normalized to $\tr \lambda^a \lambda^b=\frac{1}{2}\delta^{ab}$.

Using appropriate spinor and $SU(3)$ identities, we can rewrite this as
\bea
\mathcal L&=&\frac{g_s^2}{2}\sum_{q,q'}\left[
\gamma_{q_L,q'_L}^{ij,k\ell}\left(-\frac{1}{6}\,
\bar q_L^i\gamma^\mu q_L^j\ \bar q'^k_L\gamma^\mu q'^\ell_L+\frac{1}{2}
\bar q_L^i\gamma^\mu q'^\ell_L\ \bar q'^k_L\gamma^\mu q^j_L\right)
+L\to R\right.\nn\\
&&\phantom{x}\left.+2\,\gamma_{q_L,q'_R}^{ij,k\ell}\left(
\bar q^i_Lq_R'^\ell\ \bar q'^k_Rq^j_L-\frac{1}{3}\,
\bar q^{i\,\alpha}_Lq_R'^{\ell\,\beta} \bar q'^{k\,\beta}_R q^{j\,\alpha}_L
\right)\right]\,,
\label{gKK}
\eea
where $\alpha$ and $\beta$ are color indices~\footnote{We suppress color indices whenever they are contracted in the same way as the spinor indices.}, $q$ and $q'$ run over $u,d$ and we have defined
\be
\gamma_{q_\chi,q'_{\chi'}}^{ij\,k\ell}=y_1\int e^{2A}\left(\Omega_{q_\chi}^{ij}-\frac{y}{y_1}\delta^{ij}\right)\left(\Omega_{q'_{\chi'}}^{k\ell}-\frac{y}{y_1}\delta^{k\ell}\right)\,,
\ee
with the hermitian matrices $\Omega$ defined as
\be
\Omega_{q_\chi}^{ij}=(V_{q_\chi}\Omega^{diag}_{q_\chi} V_{q_\chi}^\dagger)^{ij} ,\quad \chi=L,R\,.
\label{Omega}
\ee
The Lagrangian (\ref{gKK}) will give rise to the leading flavor violating effects as we will see in detail in Sec.~\ref{quarkflavorgeneral}.

\subsection{Integrating out the KK-fermions}
 \label{secKKfermions}
 
We will now consider the fermion action~\cite{MertAybat:2009mk}
\begin{align}
S=& \int dy\,e^{-3A}\left(i\bar\psi_L\, /\hspace{-.22cm}\partial\,\psi_L
+i\bar\psi_R\, /\hspace{-.22cm}\partial\,\psi_R\right)\nonumber\\
&+
e^{-4A}\,\left(
\bar\psi_R\psi_L'-2A'\,\bar\psi_R\psi_L
-M_\psi(y)\bar\psi_R\psi_L+{\rm h.c.}\right)\,,
\label{A1}
\end{align}
where $\psi=(\psi_L,\psi_R)^T$ (which runs over $Q,u,d,L,\ell$) is a 5D (Dirac) fermion and for the sake of generality we have allowed the bulk mass to depend on $y$.
Defining new wave functions  
\be
\psi_{L,R}(y)=e^{2A(y)} \hat\psi_{L,R}(y)\,,
\ee
we can rewrite the Dirac equation as 
\be
m \hat \psi_{L,R}=e^{-A}(M_\psi\pm \partial_y)\hat\psi_{R,L}\,.\label{Dirac2}
\ee
For the BC we take either $\psi_L=0$ or $\psi_R=0$ at both branes.  Then there is a zero mode with profile
\be
\hat\psi^0_{L,R}(y)=\frac{e^{- Q_{\psi}(y)}}{\left(\int e^{A-2Q_{\psi}}\right)^\frac{1}{2}}\,,\qquad \hat\psi^0_{R,L}(y)\equiv 0\,,
\ee
where $Q_{\psi}(y)=\mp\int_0^y M_\psi(y')$, where the upper sign is for left handed zero modes (i.~e.~for $SU(2)$ doublets $\psi=Q,L$) and the lower one for right handed zero modes (i.e.~for $SU(2)$ singlets $\psi=u,d,\ell$).
The function $\Omega_{\psi}$ defined in Eq.~(\ref{Omega}) can then be written as
\be
\Omega_{\psi}(y)=\frac{U_{\psi}(y)}{U_{\psi}(y_1)}\,,\qquad U_{\psi}(y)=\int_0^y \exp\left[A(y')- 2Q_{\psi}(y')\right]\,.
\label{Omegafermion}
\ee
The quark Yukawa coupling is then
\be
Y_{ij}^{q}=\hat Y_{ij}^q\frac{\int h\,e^{-Q_{Q^i_L}-Q_{q^j_R}}}
  {\left(
    \int e^{-2A}h^2\int e^{A-2Q_{Q^i_L}} \int e^{A-2Q_{q^j_R}}
  \right)^{\frac{1}{2}}}\,,
\ee
where $q=(u,\, d)$. Here $\hat Y_{ij}^q$ are 5D Yukawa couplings with mass dimension $-\frac{1}{2}$.

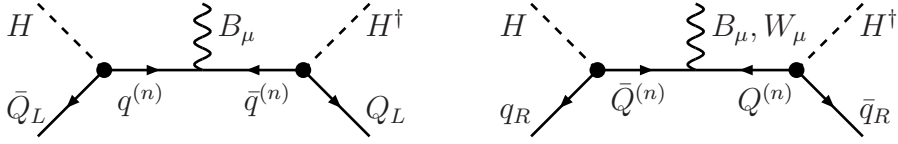
\begin{figure}[htb]
\begin{center}
\begin{picture}(65,80)(35,-20)
\SetWidth{1.}
\Vertex(25,25){3}
\Vertex(100,25){3}
\ArrowLine(25,25)(0,0)
\DashLine(0,50)(25,25){3}
\ArrowLine(25,25)(62,25)
\ArrowLine(100,25)(62,25)
\DashLine(125,50)(100,25){3}
\ArrowLine(100,25)(125,0)
\Photon(62,25)(62,50){3}{3}
\put(30,10){$ q^{(n)}$}
\put(78,10){$\bar q^{(n)}$}
\put(-12,38){$H$}
\put(124,38){$H^\dagger$}
\put(-12,7){$\bar Q_L$}
\put(124,7){$ Q_L$}
\put(68,38){$B_\mu$}
\end{picture}
%
\hspace{4cm}
\begin{picture}(65,80)(35,-20)
\SetWidth{1.}
\Vertex(25,25){3}
\Vertex(100,25){3}
\ArrowLine(25,25)(0,0)
\DashLine(0,50)(25,25){3}
\ArrowLine(25,25)(62,25)
\ArrowLine(100,25)(62,25)
\DashLine(125,50)(100,25){3}
\ArrowLine(100,25)(125,0)
\Photon(62,25)(62,50){3}{3}
\put(30,10){$\bar Q^{(n)}$}
\put(78,10){$ Q^{(n)}$}
\put(-12,38){$H$}
\put(124,38){$H^\dagger$}
\put(-12,7){$ q_R$}
\put(124,7){$\bar q_R$}
\put(68,38){$B_\mu, W_\mu$}
\end{picture}

\vspace{-1cm}
\end{center}
\caption{\it Integrating out the KK modes of the singlets (left) and doublets (right). Notice that the zero mode is not included in the internal line.}
\label{KKfermion}
\end{figure}

We would now like to integrate out the KK modes of the quarks. In particular we are interested in the diagrams shown in Fig.~\ref{KKfermion}. Notice that we can neglect these contributions if the external quarks are near UV localized. This is because, unlike the coupling of a UV localized fermion to gauge KK modes, 
the coupling of a UV localized fermion to the Higgs zero mode and a KK fermion does not go to a universal constant but rather to zero. 
Since we are primarily interested in corrections to the $Z\bar bb$ coupling we will focus on the down sector (the up sector works analogously with $H\to i\sigma_2 H^*$). One obtains the effective 4D Lagrangian of zero modes
\bea
 \mathcal L'_{\rm non-oblique} &=&i \beta^{d_L}_{i\ell}\, \bigl[\bar Q_L^i\,H\bigr]\, / \!\!\!\! D\,\bigl[ H^\dagger\, Q^\ell_L\bigr]
+i\beta^{d_R}_{i\ell}
\, \bigl[\bar d_R^i\,H^\dagger\bigr]\, / \!\!\!\! D\,\bigl[ H\, d^\ell_R\bigr]
\nn\\
&\supset&\beta^{d_L}_{i\ell}\frac{v^2}{4}\frac{g}{c_w}\bar d_L^i \gamma^\mu Z_\mu\,d_L^\ell
-\beta^{d_R}_{i\ell}\frac{v^2}{4}\frac{g}{c_w}\bar d_R^i \gamma^\mu Z_\mu\,d_R^\ell
\, ,
\label{Zbb2}
\eea
where in the second line we have also used the Dirac equation. 
Using the results of App.~\ref{app:fermions} we can express the couplings $\beta$ as
\bea
\beta^{d_L}_{i\ell}&=&
\sum_jY_{ij}^{d}Y_{\ell j}^{d*}
\int_0^{y_1}
e^{2A}
\,
(\Omega'_{d_R^j})^{-1}
(\Gamma^d_{\ell j}-\Omega_{d^j_R})(\Gamma^d_{ij}-\Omega_{d_R^j})\,,\nn\\
\beta^{d_R}_{i\ell}&=&
\sum_jY_{ji}^{d^*}Y_{j\ell}^{d}
\int_0^{y_1}
e^{2A}
\,
(\Omega'_{d_L^j})^{-1}
(\Gamma^d_{j\ell}-\Omega_{d^j_L})(\Gamma^d_{ji}-\Omega_{d_L^j})\,,\label{beta}
\eea
where we have defined
\be
\Gamma^d_{ij}(y)=\frac{\int_0^{y}h\,e^{-Q_{Q_L^i}-Q_{d_R^j}}}{\int_0^{y_1}h\,e^{-Q_{Q_L^i}-Q_{d_R^j}}}\,,
\ee
which is the cumulative distribution of the physical 4D down-type Yukawa coupling along the fifth dimension. Thus, $\Gamma$ monotonically increases from zero to one; if any of the three fields (Higgs, $Q_L^i$ or $d_R^j$) is UV (IR) localized we can take the limit $\Gamma^d_{ij}\to 1$ ($\Gamma^d_{ij}\to 0$)~\footnote{Using this simple limit we have checked that we obtain the same result as quoted in Ref.~\cite{Casagrande:2008hr} for an IR-brane localized Higgs.}. 

The non-oblique Lagrangian (\ref{Zbb2}) will contribute with a significant amount to the $Z\bar bb$ coupling as we will describe in detail in Sec.~\ref{EWPT}. 

\section{Quark masses and mixing angles}
\label{quarks}

\begin{figure}[htb]
\begin{center}
\begin{psfrags}
\input{F-psfrag.tex}
\includegraphics[width=0.5\textwidth]{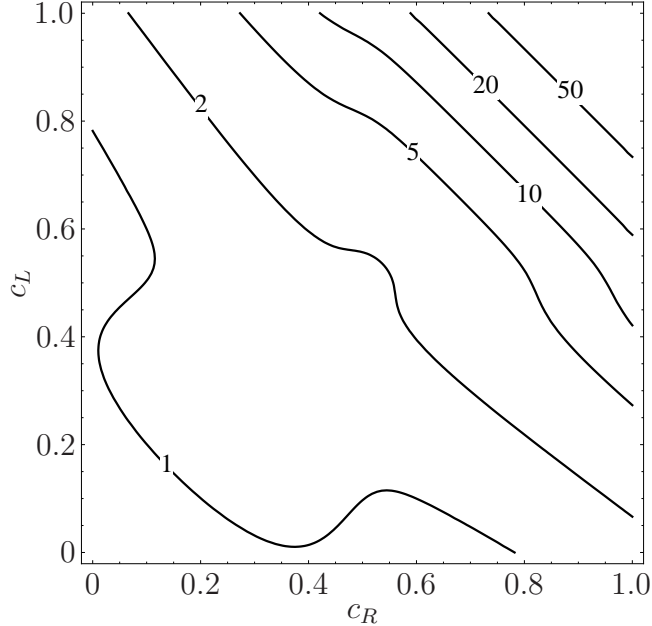}
\end{psfrags}
\end{center}
\caption{\it The function $F(c_L,c_R)$ for $k\Delta=1,\ \nu=0.5$ normalized to the corresponding RS value. One can see that for a wide portion of the parameter space it leads to an enhancement of the 4D Yukawa couplings with respect to RS. 
}
\label{yukawaenhancement}
\end{figure}

In this section we will introduce explicit quark zero mode profiles and fit the parameters in the 5D Lagrangian to the observed quark masses, mixing angles and $CP$-violating phase. We will make the choice 
\begin{equation}
Q_\psi(y)=c_\psi A(y)\,,
\label{choice}
\end{equation}
which concides with that used in RS models where $Q^{RS}_\psi=c_\psi ky$~\footnote{Of course one can also use for a general model $Q_\psi=c_\psi ky$. See e.g.~Ref.~\cite{Carmona:2011ib}.}. 
This particular mass term can be achieved if the stabilizing field $\phi$ couples to the fermions. In particular if we parametrize the bulk potential for $\phi$ by a "superpotential" $W(\phi)$ as $V=3W'^2-12W^2$~\cite{DeWolfe:1999cp} we can achieve Eq.~(\ref{choice}) by the choice $M_\psi(\phi)=\mp c_\psi W(\phi)$ where the upper sign holds for 5D fermions with left-handed zero modes (i.~e.~$\psi=Q_L,L_L$) and the lower one for 5D fermions with right-handed zero modes (i.~e.~$\psi=u_R,d_R,\ell_R$). In this case the previous definitions simply read
\begin{equation}
U_\psi(y)=\int_0^y \exp\left[(1-2c_\psi)A(y')\right]\,,
\end{equation}
and
\be
Y_{ij}^q=\hat Y_{ij}^q \ F(c_{Q^i_L},c_{q^j_R})\,,
\ee
where $q=(u,d)$ and the function $F$ is defined as
\begin{equation}
F(c_L,c_R)=\frac{\int h\,e^{-(c_{L}+c_{R})A}}
  {\left[
    \int e^{-2A}h^2\int e^{ (1-2 c_{L})   A} \int e^{(1-2 c_{R})A}
  \right]^{\frac{1}{2}}}\ .
\end{equation}
We note the following properties of the fermion profiles and the function $F$.
\begin{itemize}
\item
For any strength of the metric deformation, fermions $\psi$ are IR (UV) localized for $c_\psi<\frac{1}{2}$ ($c_\psi>\frac{1}{2}$). This is thus the same situation as in the RS model. Notice also that this choice of profile corresponds, in the dual theory, to the special case of constant anomalous dimension, i.e.~the fermionic operators are not significantly disturbed by the presence of the CFT deformation.
\item
The larger the deformation of the AdS background the larger is the portion of the parameter space $(c_L, c_R)$ where the function $F(c_L,c_R)$ is enhanced. Consequently,
 the coefficients $c_\psi$  can be slightly larger for the same 5D Yukawa coupling in order to reproduce the same (physical) 4D Yukawa coupling. This in turn leads to a weaker coupling of the fermions to the KK modes of the gauge fields. Alternatively, for fixed values of the coefficients $c_\psi$ and 4D Yukawa couplings, the 5D Yukawa couplings and correspondingly the couplings of fermion KK-modes to the fermion and Higgs zero modes in Fig.~\ref{KKfermion} are decreased with respect to their values in the AdS background leading to a softening of the bounds on the value of $m_{KK}$ as we will see in Sec.~\ref{EWPT}. This enhancement of the function $F(c_L,c_R)$ for a background with a large AdS deformation is illustrated in Fig.~\ref{yukawaenhancement} for the metric given in Eq.~(\ref{ourmetric}) with $k\Delta=1$ and $\nu=0.5$.
\end{itemize}

Let us now consider the quark mass-squared matrices
\be
\M^{q_L}=\frac{v^2}{2}\,  Y^{q} Y^{q\,\dagger},\qquad q=u,d
\ee
where $v=246$ GeV is the VEV of the Higgs field. Unitary matrices should be introduced to diagonalize the matrices $\M^{q_L}$ as
\be
\M_{diag}^q=V^{q_L}\,\M^{q_L}\,V^{q_L\,\dagger}\,.
\label{rotaciones}
\ee
Next let us write expressions for the eigenvalues and mixing angles of the hierarchical Yukawa couplings. In the following we will just make two reasonable assumptions
\begin{itemize}
\item
First we will assume a left handed hierarchy, i.e.
\be
Y^q_{1i}\ll Y^q_{2i}\ll Y^q_{3i}.
\label{LHhierarchy}
\ee
This will be the case whenever there is a mild hierarchy between the left-handed $c_\psi$, i.e.~$c_{Q_L^1}>c_{Q_L^2}>c_{Q_L^3}$~\footnote{We will comment later on about possible further simplifications that take place in case there is also a right handed hierarchy.}.
\item
The only second assumption we are making is that the left handed rotations show a similar hierarchy as the CKM matrix
\be
|V^{q_L}_{12}|\sim \epsilon,\quad |V^{q_L}_{23}|\sim \epsilon^2\,,\quad |V^{q_L}_{13}|\sim \epsilon^3\,,
\label{rotations}
\ee
where $\epsilon$ is of the order of the Cabbibo angle. 
This assumption is natural since otherwise the smallness of the CKM mixing angles would be a consequence of large cancellations. As it turns out it is also consistent with the assumption Eq.~(\ref{LHhierarchy}).
\end{itemize}

Owing to our assumption Eq.~(\ref{rotations}) these unitary rotations can be given in a Wolfenstein-like parameterization as~\cite{PDG}
\be
V^{q_L}=\left(
\begin{array}{ccc}
1-\frac{1}{2}|V^{q_L}_{12}|^2&V^{q_L}_{12}&V^{q_L}_{13}\\
-V^{{q_L}*}_{12}&1-\frac{1}{2}|V^{q_L}_{12}|^2&V^{q_L}_{23}\\
(-V^{{q_L}}_{13}+V^{q_L}_{12}V^{q_L}_{23})^*&-V_{23}^{{q_L}*}&1
\end{array}
\right)\,,\qquad q=u,d
\label{matrices}
\ee
where terms of order $\mathcal O(\epsilon^4)$ have been neglected. The matrix $V^{q_L}$ contains three independent complex parameters and it is unitary to the considered order. The angles and eigenvalues are best expressed in terms of the quantities
\be
\widetilde \M^{q_L}_{ij}=\M^{q_L}_{ij}-\frac{\M^{q_L}_{i3}\M^{q_L}_{3j}}{\M_{33}}\,.
\ee
First, by demanding the off-diagonal terms in Eq.~(\ref{rotaciones}) to vanish we obtain the mixing angles
\begin{align}
V^{q_L}_{12}&=-\widetilde \M^{q_L}_{12}\,/\,\widetilde \M^{q_L}_{22}\,,\qquad\qquad
V^{q_L}_{21}=\widetilde \M^{q_L}_{21}\,/\,\widetilde \M^{q_L}_{22}\,,\nn\\
V^{q_L}_{23}&=-\M^{q_L}_{23}\,/\,\M^{q_L}_{33}\,,\qquad\qquad
V^{q_L}_{32}=\M^{q_L}_{32}\,/\,\M^{q_L}_{33}\,,\nn\\
V^{q_L}_{13}&=-\M^{q_L}_{13}\,/\,\M^{q_L}_{33}
+(\widetilde\M^{q_L}_{12}\M^{q_L}_{23})/(\widetilde\M^{q_L}_{22}\M^{q_L}_{33})\,,\nn\\
V^{q_L}_{31}&=\M^{q_L}_{31}\,/\,\M^{q_L}_{33}\,,
\label{angulos}
\end{align}
The mass eigenvalues are then obtained as:
\bea
(m_3^q)^2&=&\M_{33}^{q_L},\nonumber\\
(m_2^q)^2&=&\widetilde \M_{22}^{q_L}\,,\nn\\
(m_1^q)^2&=&\widetilde\M_{11}^{q_L}-\frac{\widetilde\M_{12}^{q_L}\widetilde\M_{21}^{q_L}}{\widetilde\M^{q_L}_{22}}\,,
\label{masascomp}
\eea
where we are using the notation $m_3^u=m_t$, $m_3^d=m_b$, and so on.
The comparison with CKM matrix ($V=V^{u_L} V^{d_L\dagger}$) elements leads to the relations
\bea
V_{us}&=&
   \hat V_{12}\,,\nonumber\\
V_{cb}&=&
   \hat V_{23}\,,\nonumber\\
V_{ub}&=&
   (-\hat V_{31}+V_{21}^{u_L}\hat V_{32})^*\,,\nonumber\\
V_{td}&=&
   \hat V_{31}-V_{21}^{d_L}\hat V_{32}\,,
\label{CKMcomp}
\eea
where $\hat V=V^{u_L}-V^{d_L}$.
The CKM matrix defined this way does not obey the usual phase convention~\cite{PDG}. One can easily obtain the standard convention  by multiplying $V^{q_L}$ from the left with appropriate phases. One finds
\be
e^{i\delta}=\frac{V^*_{ub}V_{us} V_{cb}}{|V_{ub}V_{us}V_{cb}|}\,.
\ee
Alternatively we can write the Jarlskog invariant as
\bea
J&=&\operatorname{Im}\,( V^*_{ub}V^*_{sc}V_{us} V_{cb})\nn\\
   &=&-\operatorname{Im}\,(\hat V_{12}\hat V_{23}\hat V_{31})+|V_{cb}|^2
    \operatorname{Im}\,(V_{12}^{d_L}V^{u_L}_{21})\,.
   \label{Jarlskog}
\eea

To summarize there are nine constants $c_{Q^i_L},c_{u^i_R},c_{d^i_R}$ which should be adjusted to satisfy the mass relations (\ref{masascomp}) and the experimental CKM matrix relations (\ref{CKMcomp}) and (\ref{Jarlskog}).
We have performed a $\chi^2$ fit to the experimental quark masses (measured at the KK scale~\cite{Csaki:2008zd}), the mixing angles and the phase. To this end we have 
randomly generated a set of 40,000 complex 5D Yukawa couplings and fitted to the nine parameters $c_\psi$ such that we correctly reproduce the experimental data.
We have accepted points which yield a $\chi^2 \lesssim 4$ to {\em both} a pure RS metric and to metric (\ref{ourmetric}) with $\nu=0.5,\ k\Delta=1$. Each pair of 5D Yukawas ($\hat Y^u_{ij}$ and $\hat Y^d_{ij}$) thus gives rise to two sets of $c_\psi$, one for RS and one for the model defined by Eq.~(\ref{ourmetric}). This facilitates a direct comparison between the two models since the 4D effective theories originate from the same set of 5D Yukawa couplings.
We have chosen flat prior distributions  $1\leq |\sqrt{k}\hat Y^q_{ij}|\leq 4$ and $0\leq \arg(\hat Y^q_{ij})<2\pi$.

\begin{figure}[p!]
\centering
\input{cPDF-psfrag.tex}
\includegraphics[width=0.85\textwidth]{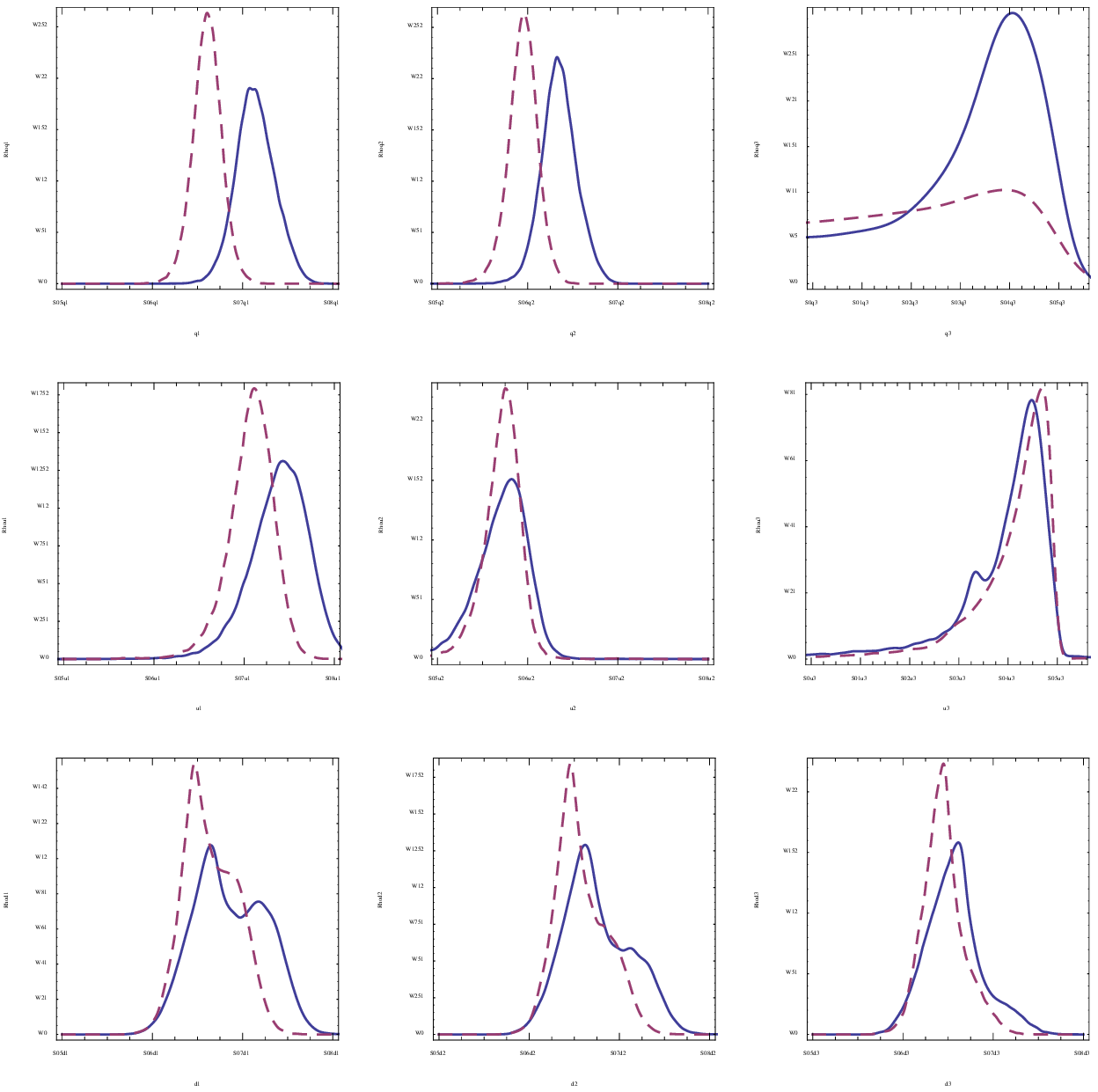}
\caption{\it The distribution of the $c$ parameters for the different 
quarks and chiralities, for RS and for model (\ref{ourmetric}) with $k\Delta=1$ and $\nu=0.5$. Only the parameters $c_{(t,b)_L}$ and $c_{t_R}$ are IR localized ($c<0.5$). Notice the highly assymetric forms of the corresponding distributions. } 
\label{figc}
\end{figure}
\begin{table}[p!]
\centering
RS:~
\begin{tabular}{| l | l | l |}
\hline
$c_{(u,d)L} =0.66 \pm 0.02$ & $c_{(c,s)L} = 0.59 \pm 0.02  $ & $c_{(t,b)L} =  -0.11^{+0.45}_{-0.53} $
\\ \hline 
$c_{uR} = 0.71 \pm 0.02$ & $c_{cR} = 0.57 \pm 0.02$ & $c_{tR} =  0.42^{+0.05}_{-0.17}$
\\ \hline
$c_{dR} = 0.66 \pm 0.03$ & $c_{sR} = 0.65 \pm 0.03 $ & $c_{bR} = 0.64 \pm 0.02$ 
\\ \hline
\end{tabular}
\vspace{10pt}
\hspace{-122pt}
 $
 \nu=0.5$:~
\begin{tabular}{| l | l | l |}
\hline
$c_{(u,d)L} = 0.71 \pm 0.02 $ & $c_{(c,s)L} =  0.63 \pm 0.02$ & $c_{(t,b)L} =  0.31^{+0.13}_{-0.52}$
\\ \hline
$c_{uR} =  0.74 \pm 0.03$ & $c_{cR} =  0.57 \pm 0.03$ & $c_{tR} =  0.42^{+0.05}_{-0.11}$
\\ \hline
$c_{dR} = 0.68 \pm 0.04$ & $c_{sR} = 0.67 \pm 0.04$ & $c_{bR} =  0.66 \pm 0.03$
\\ \hline 
\end{tabular}

~

\caption{\it Medians and $1\sigma$ confidence intervals of the $c$ parameters corresponding to the different species of quarks and chiralities, for RS and for model (\ref{ourmetric}) with $k\Delta=1,\,\nu=0.5$. }
\label{tabc}
\end{table}
The results of the fit are presented in Fig.~\ref{figc} while the corresponding central values and $1\sigma$ confidence intervals of the $c_\psi$ parameters are listed in Tab.~\ref{tabc}.
As it is clear from the individual plots, the  $c_{\psi}$ are slightly larger for our model than in the RS model, as anticipated above~\footnote{We have restricted ourselves to the region $c_\psi>-1$ in order to avoid strongly IR localized fermions, which typically have stricter perturbativity bounds for the Yukawa couplings. }. 
This means that the couplings of the individual quarks to KK modes are more suppressed than in RS. \footnote{We have checked that for fixed $c$, the individual couplings of KK gauge bosons to fermion zero modes are of the same order for both models.}

An interesting fact that we find is that the $c_{d^i_R}$ are very much non-hierarchical. In fact only about 30\% of all points show the "traditional" hierarchy $c_{d_R^1}>c_{d_R^2}>c_{d_R^3}$. 
As expected our expressions Eq.~(\ref{angulos}) and Eq.~(\ref{masascomp}) are much better approximations to the true angles and eigenvalues in these cases than the ones usually employed in the literature~\cite{Hall}. Note that in practice we never need to have explicit expressions for the right handed angles in terms of the Yukawa matrices, as the former do not enter in the fit~\footnote{For each data point obtained in the fit it is of course a simple matter to numerically find the right handed rotations.}.
On the other hand, the up-type sector will always be hierarchical, $c_{u_R^1}>c_{u_R^2}>c_{u_R^3}$ or 
$Y^u_{i1}\ll Y^u_{i2}\ll Y^u_{i3}$ respectively, and we could have used the simpler expressions for the eigenvalues and angles described in App.~\ref{RH}.

\section{Electroweak Precision Tests}
\label{EWPT}

The general procedure to evaluate the effect oblique and non-oblique EWPO's in the presence of New Physics is briefly described in App.~\ref{EWPO} where extensive use has been done of Ref.~\cite{Burgess:1993vc}. In this section we just present the final results in general warped spaces which are ready for numerical calculations in particular models, as models with RS metric and the metric in Eq.~(\ref{ourmetric}). 

The oblique Lagrangian Eq.~(\ref{oblique}) generates the $(T,\, S,\, W,\, Y)$  parameters as given in Ref.~\cite{Cabrer:2010si}
\bea
\alpha T&=&s_w^2m_Z^2\,\hat\alpha_{hh}\,,\nn\\
\alpha S&=&8s_w^2c_w^2m_Z^2\,\hat\alpha_{hg}\,,\nn\\
Y&=&c_w^2m_Z^2\,\hat\alpha_{gg}\,,\nn\\
W&=&Y\,.
\eea
Let us now focus on the non-oblique operators. In particular, 
 the operator coupling the Higgs current to the fermion currents -- the first operator in 
Eq.~(\ref{non-oblique1}) -- contributes to the modified $Z$ and $W$ vertices. 
The same holds true for the operator given in Eq.~(\ref{Zbb2}).
It is straightforward to extract the contributions to the vertex corrections by going to the electroweak vacuum~\footnote{
Notice, in particular that from Eq.~(\ref{eom}) one has
$
\Box A_\mu^A = -J_{h\,\mu}^A+\dots.
$
Hence, after EWSB, $ J_h=-\mathcal M^2 A$ where $\mathcal M^2$ is the gauge boson mass matrix and one can directly evaluate the product with the fermion currents.}.
Diagonalizing the physical Yukawa couplings with the rotation matrices $V_{d_\chi}$, the $Z\bar qq$ couplings receive the corrections 
\bea
\delta g_{q^{1,2}_{L,R}}&=&\frac{g^{SM}_{q_{L,R}}}{2} \left(\alpha T+\frac{Y}{c_w^2}\right)
-Q^{em}_{q}\,\frac{1}{c_w^2-s_w^2}\left(\frac{\alpha S}{4}-c_w^2 s_w^2\,\alpha T-s_w^2\,Y\right)\,,\quad (q=u,d)\nn\\
\delta g_{b_{L,R}}&=&
\frac{g^{SM}_{d_{L,R}}}{2} \left(\alpha T+\frac{Y}{c_w^2}\right)
+\frac{1}{3}\,\frac{1}{c_w^2-s_w^2}\left(\frac{\alpha S}{4}-c_w^2 s_w^2\,\alpha T-s_w^2\,Y\right)
+\delta\tilde g_{b_{L,R}}\,,\nn\\
\delta\tilde g_{b_L}&\equiv&\delta\tilde g^{33}_{d_L}=\left(-g^{SM}_{d_{L}}\, m_Z^2\, 
\hat\alpha_{h,d^i_{L}} \delta_{i\ell}+\frac{v^2}{4}\beta^{d_L}_{i\ell}\right)
V^{3i}_{d_L}V_{d_L}^{*3 \ell}\,,\nn\\
\delta \tilde g_{b_R}&\equiv&\delta\tilde g^{33}_{d_R}=\left(-g^{SM}_{d_{R}}\, m_Z^2\, 
\hat\alpha_{h,d^i_{R}} \delta_{i\ell}-\frac{v^2}{4}\beta^{d_R}_{i\ell}\right)
V^{3i}_{d_R}V_{d_R}^{*3 \ell}\,,
\label{deltagoverg}
\eea
where $g_{q_L}^{SM}=T^3_{q}-Q^{em}_q\,s_w^2$ and $g^{SM}_{q_R}=-Q^{em}_qs_w^2$, and the integrals $\hat\alpha_{h\psi}$ and $\beta_{ij}^\psi$ have been given in Eqs.~(\ref{alphahat}) and (\ref{beta}).
The tilde here denotes an explicit vertex correction coming from the non-oblique operators. The dependence of the couplings on the oblique parameters result from the various effects mentioned in Ref.~\cite{Burgess:1993vc} and App.~\ref{EWPO}.
We have already mentioned that we can neglect the diagrams in Fig.~\ref{KKfermion} if the external quarks are near UV localized. Moreover, the contribution from the gauge KK modes is universal for near UV localized modes and is summarized in the oblique parameters. As shown in Sec.~\ref{quarks} only the left handed top-bottom doublet and the right handed top singlet are near IR localized. We will thus neglect the explicit correction $\delta \tilde g_{b_R}$.

The analysis for oblique observables was already performed for general models in Refs.~\cite{Cabrer:2010si,Cabrer:2011vu,Carmona:2011ib} and the resulting bounds do depend to a large extent on the value of the  mass of the Higgs boson and its location along the fifth dimension.  The general result is that the less localized the Higgs  towards the IR brane [the lower the value of the $a$ parameter in Eq.~(\ref{ourmetric})] the milder bounds oblique corrections impose on the mass of KK modes. On the other hand the degree of delocalization of the Higgs is bounded by the solution of the hierarchy problem which imposes a lower bound on the $a$ parameter. For instance for the RS model the solution of the hierarchy problem imposes the bound $a\gtrsim 2$ and for a light Higgs $m_H\simeq 115$ GeV oblique observables impose on the KK modes the bound $m_{KK}\gtrsim 7.5$ TeV~\footnote{Notice that this bound can be lowered to $\sim 3-4$ TeV by introducing an extra gauge custodial symmetry in the bulk~\cite{Agashe:2003zs}.}. Moreover for the model of Eq.~(\ref{ourmetric}) with $k\Delta=1$ and $\nu=0.5$ solving the hierarchy problem imposes the bound $a\gtrsim 3.1$ and for $m_H\simeq 115$ GeV the oblique observables are consistent with experimental data for $m_{KK}\gtrsim 1$ TeV. The reason for this improvement is that in the deformed background the Higgs zero mode can become more decoupled from the KK modes than in the pure RS model.

\begin{figure}[htb]
\begin{center}
\includegraphics[width=0.48\textwidth]{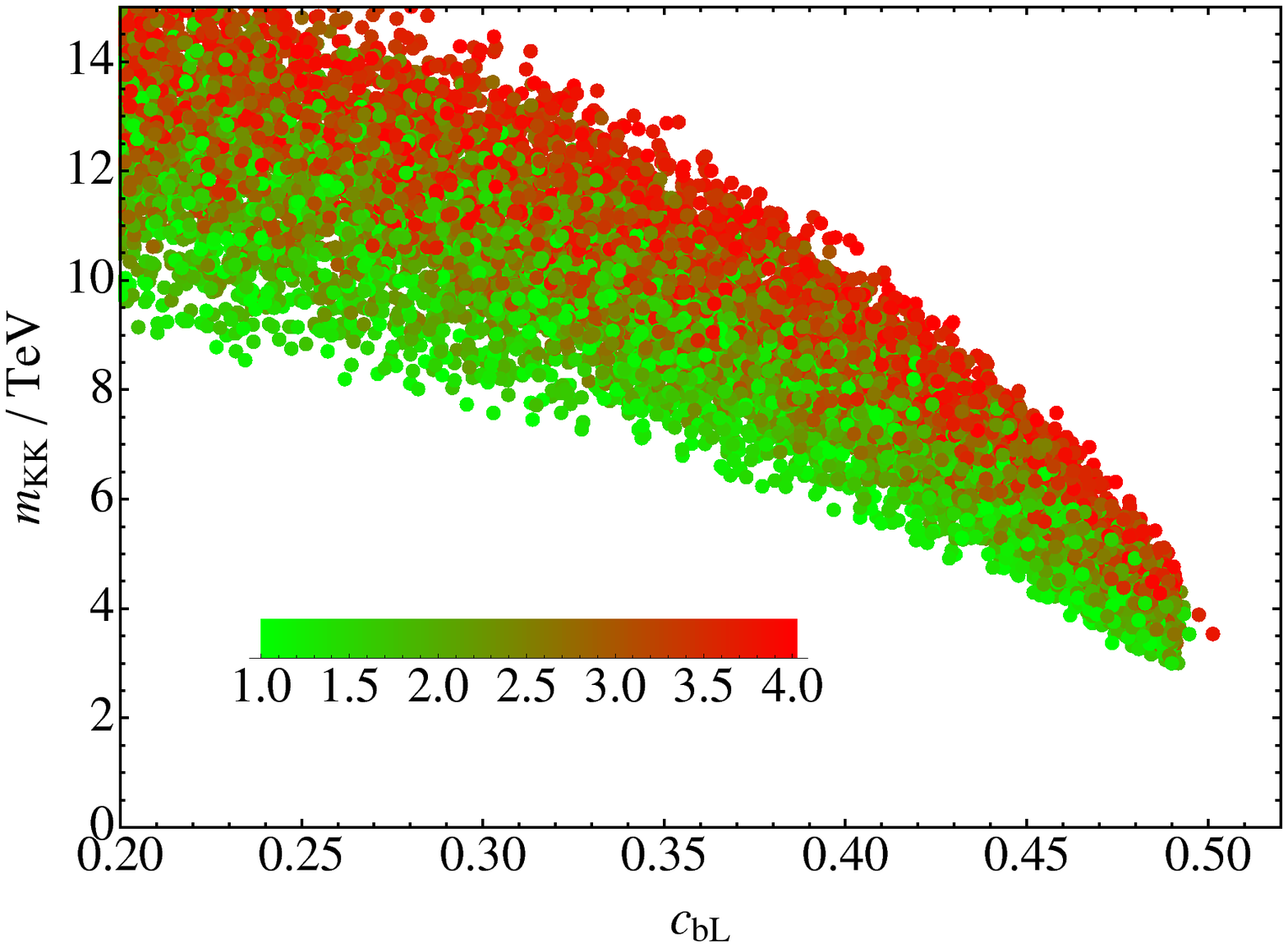}
~
\includegraphics[width=0.48\textwidth]{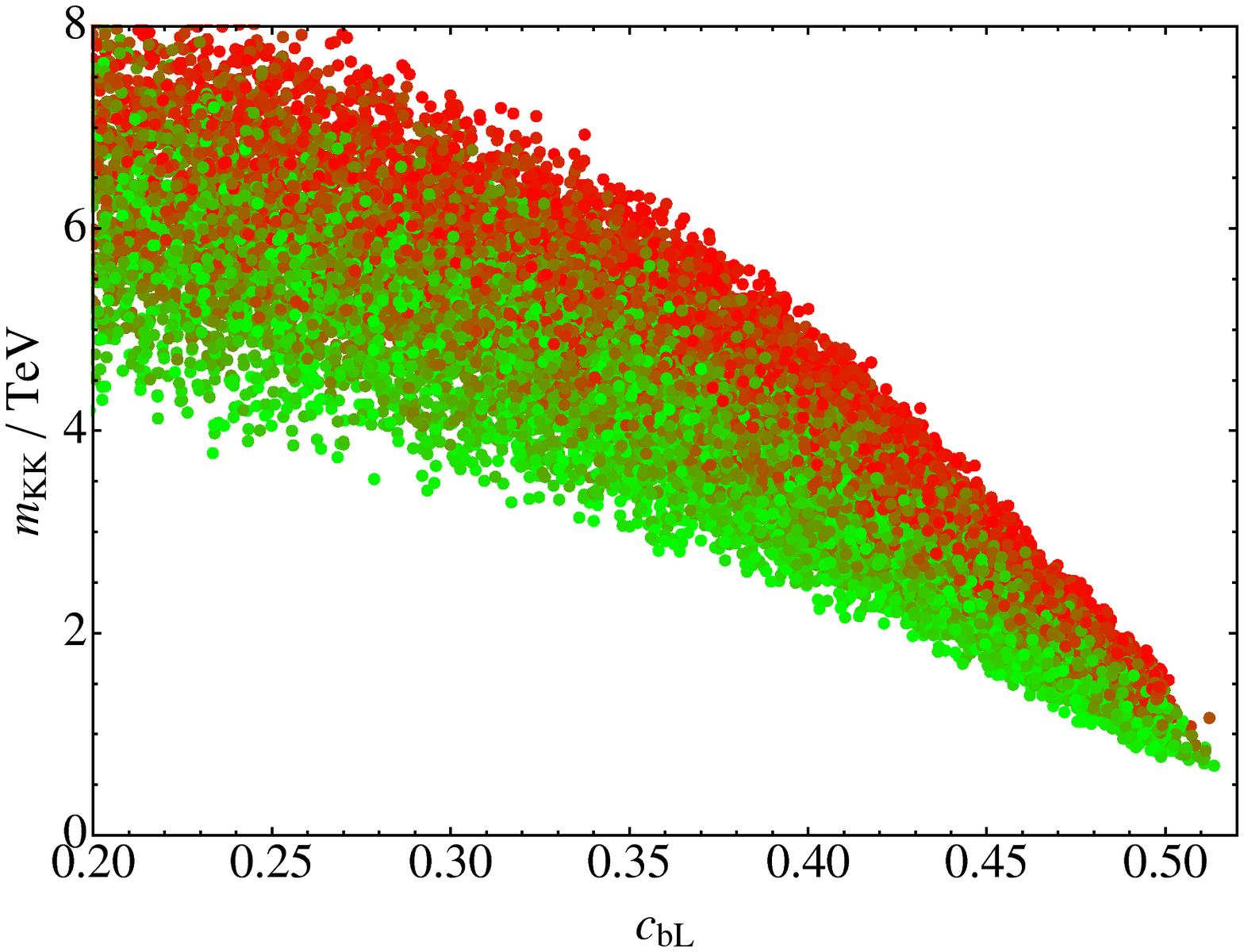}
\end{center}
\caption{\it The bounds (in TeV) implied by the experimental limits on $R_b$, as a function of $c_{Q_L^3}$. Left panel: RS model. Right panel: our model with $k\Delta=1$, $\nu=0.5$. We also display the dependency on the 5D bottom Yukawa:  the coloring interpolates between green (light gray) for $\sqrt{k}\hat Y^d_{33}=1$ to red (dark gray) for $\sqrt{k}\hat Y^d_{33}= 4$.}
\label{figZbb2}
\end{figure}

There are several factors which influence the size of the nonuniversal $Z\bar bb$ coupling.
\begin{itemize}
\item
The more UV localized the left handed top-bottom doublet, the more suppressed are its coupling to the KK modes of the gauge bosons and to those of the singlet quarks. Hence we expect a suppression of the contribution to $Z\bar b b$ for larger $c_{Q_L^3}$. 
\item
The smaller the 5D bottom Yukawa $\hat Y^d_{33}$, the more suppressed the Yukawa coupling of $b_L$ to the singlet KK modes appearing in the left panel of Fig.~\ref{KKfermion}, and hence the more suppressed is this contribution to $\delta \tilde g_{b_L}$.
\item
As explained above, the Higgs can become more decoupled from the IR in the deformed background, and this reduces both the coupling to KK gauge bosons and KK fermions. 
\end{itemize} 

In order to compute the effect of the nonuniversal $Z\bar bb$ coupling to the partial width
\be
R_b=\frac{\Gamma(Z\to \bar b b)}{\Gamma(Z\to \bar q q)}
\ee
we write
\be
R_b=R_b^{SM}+\left.\left(\sum_{q\neq t}\frac{\partial R_b^{tree}}{\partial g_{q_\chi}}\delta g_{q_\chi}\right)\right|_{g_{q_\chi}^{SM}}\,,
\ee
where 
\be
R_b^{tree}=\frac{g_{b_L}^2+g_{b_R}^2}{\sum_{q\neq t}(g_{q_L}^2+g_{q_R}^2)}\,,\qquad
R_b^{SM}=0.21578\,.
\ee
Only the left handed bottom has both oblique and non-oblique corrections, while all other couplings only have corrections coming from the oblique parameters, see Eq.~(\ref{deltagoverg}). The result should be compared to the experimental value \cite{PDG}
\be
R_b^{exp}=0.21629\pm0.00066\,,
\ee
and translates into a lower bound on the KK scale.

All effects enumerated above are clearly visible in Fig.~\ref{figZbb2} where we present plots of the minimal KK scale required to sufficiently suppress the observable
 $R_b$ as computed in Eq.~(\ref{deltagoverg}). 
We have used the randomly generated set of data used to fit the quark masses, mixing angles and $CP$ violating phase in Fig.~\ref{figc}.
In particular, the third effect above reduces the bounds (for fixed $c_{Q_L^3}$ and $\hat Y^d_{33}$) by roughly a factor of 2 when comparing the RS model to the model defined by Eq.~(\ref{ourmetric}) for $k\Delta=1$ and $\nu=0.5$. We have also checked the dependence on the choice of the fermion bulk mass term. In particular Ref.~\cite{Carmona:2011ib} used a constant mass term $M(y)=c k$. Although our analysis is quite different, we have verified the results in Ref.~\cite{Carmona:2011ib} qualitatively. In particular, for the anarchic case the bounds are slightly higher than the ones for the choice $M(y)=cA'(y)$ and show a stronger dependence on the 5D bottom Yukawa coupling. This indicates that the effect of the KK fermions is dominating for large $\hat Y^d_{33}$, which can easily be mitigated by lowering that coupling at the cost of a mild $\mathcal O(10)$ hierarchy  in the 5D Yukawa couplings \cite{Carmona:2011ib}.

\begin{figure}[htb]
\begin{center}
\begin{psfrags}
\input{ZbbPDF-psfrag.tex}
\includegraphics[width=0.48\textwidth]{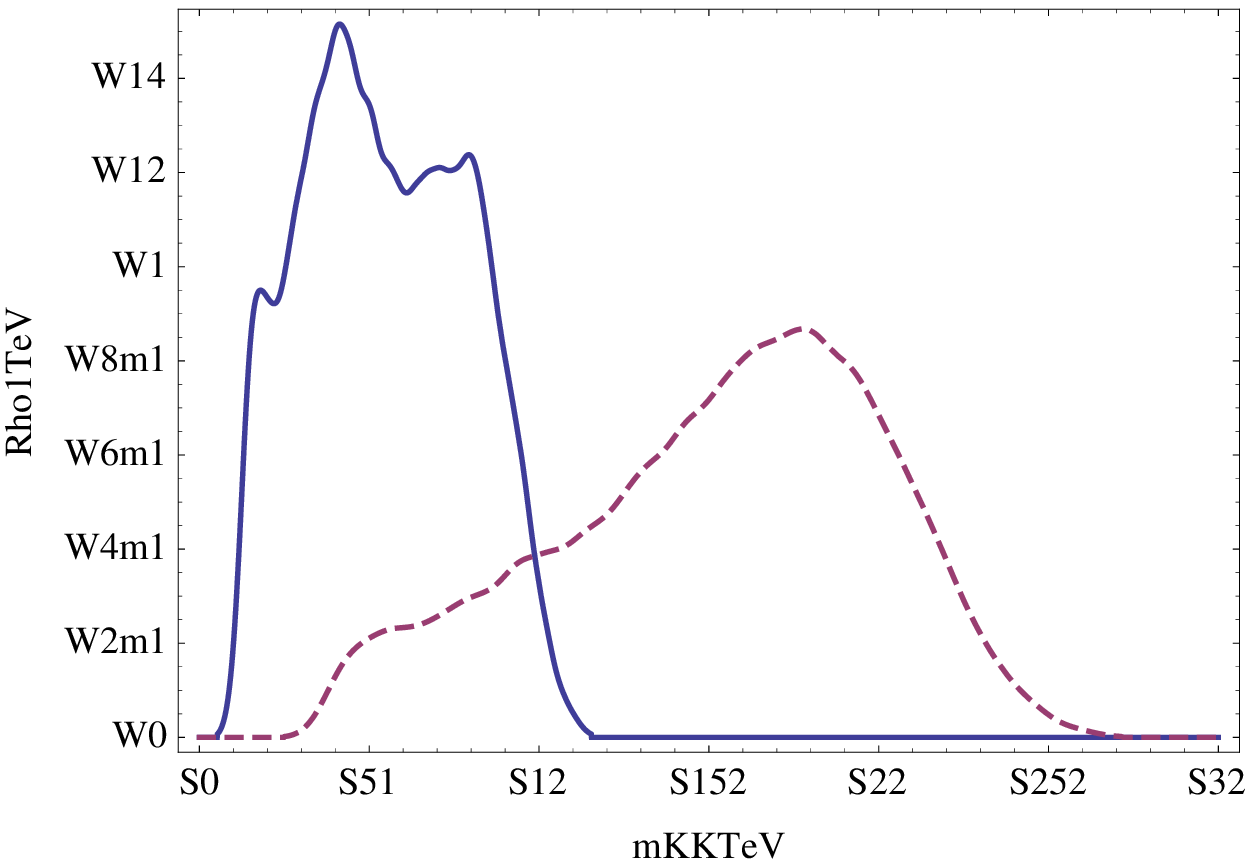}
\end{psfrags}
~
\begin{psfrags}
\input{ZbbCDF-psfrag.tex}
\includegraphics[width=0.48\textwidth]{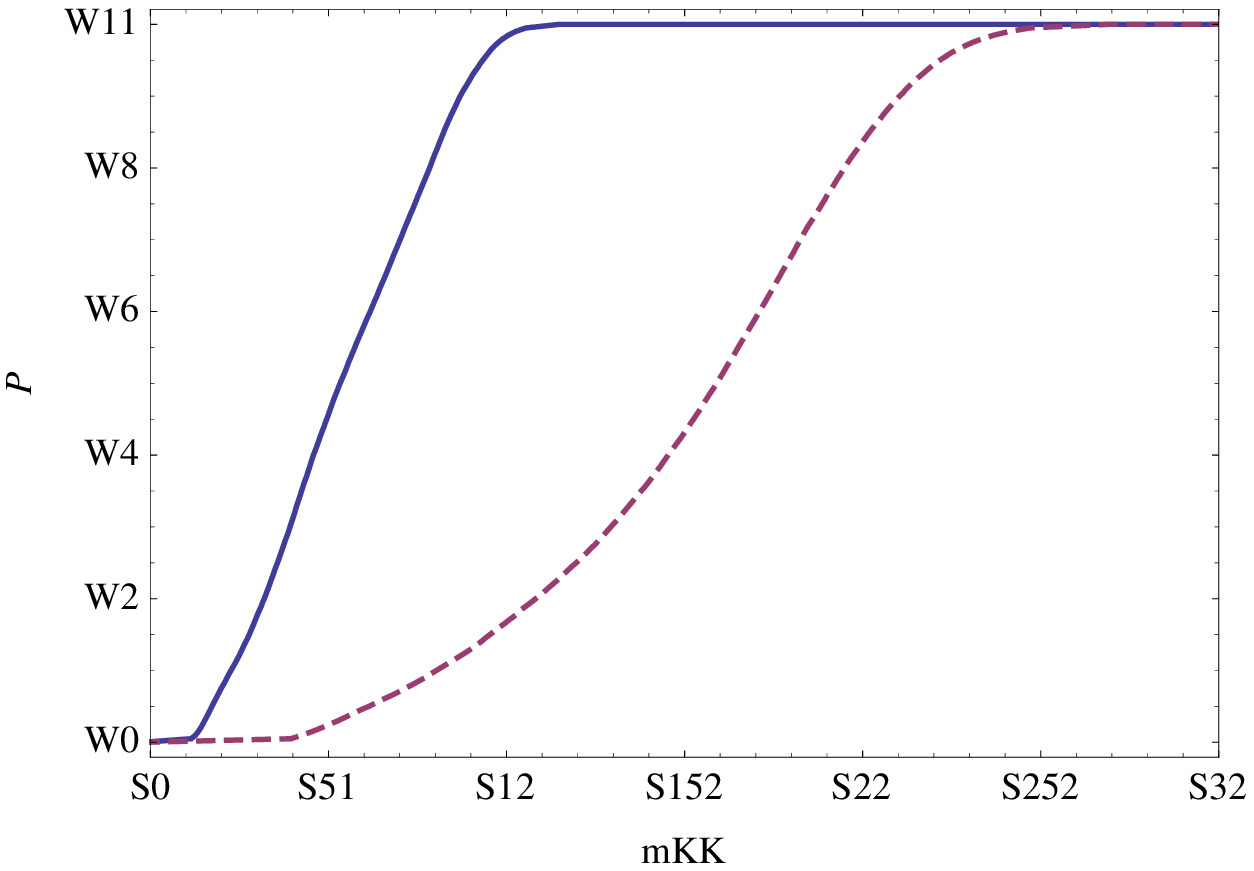}
\end{psfrags}
\end{center}
\caption{\it PDF (left panel) and CDF (right panel) for $m_{KK}$ from comparison with $R_b$. Dashed lines correspond to the RS model and solid lines to the model in Eq.~(\ref{ourmetric}) for $k\Delta=1$ and $\nu=0.5$.}
\label{figZbb}
\end{figure}

Moreover in Fig.~\ref{figZbb}  we have considered the probability distribution functions (PDF) and cumulative distribution functions (CDF) for the  lower bound on $m_{KK}$. The fact that the model with the modified background generally requires larger $c_{Q^3_L}$, see Fig.~\ref{figc}, further pushes these distributions to lower KK scales, implying milder bounds.
For a given KK scale on the horizontal axis one can read off from the CDF (right panel of Fig.~\ref{figZbb}) the fraction of points consistent with such a scale for both models on
the vertical axis. This fraction is thus the probability that the KK scale is smaller than a given value.
Notice that it can also be viewed as the amount of fine tuning necessary to obtain a given bound. Conversely one can start from a given fraction (fine tuning) and read off the percentile on the horizontal axis for both models. 
Moreover in Tab.~\ref{tabZbb} we present some explicit numbers obtained from these distributions. 
\begin{table}[htb]
\centering
\begin{tabular}{|c|ccc|ccc|}
\cline{2-7}
\multicolumn{1}{c|}{}&\multicolumn{3}{c|}{\textit{Probability for $m_{KK}$ below}}&\multicolumn{3}{c|}{\textit{Percentile}}\\ 
\multicolumn{1}{c|}{}&$ 3$ TeV &$5$ TeV &$10$ TeV&10\% &20\% &50\% \\
\hline
RS &0\% 			&2.4\% &17\% &8.0 TeV&11 TeV& 16 TeV\\
$\nu=0.5$ &18\% 	&46\% &98\% &2.3 TeV&3.2 TeV& 5.3 TeV\\
\hline
\end{tabular}
\caption{\it Left panel: Integrated probability for values of $m_{KK}$ below 3, 5 and 10 TeV from $R_b$ for RS (upper row) and the model in Eq.~(\ref{ourmetric}) for $k\Delta=1$ and $\nu=0.5$ (lower row). Right panel: 10th, 20th and 50th (median) percentiles for both models.}
\label{tabZbb}
\end{table}
As we can see from Tab.~\ref{tabZbb} getting "acceptable" bounds depends to a large extent on the amount of fine-tuning which we tolerate. For instance assuming a 20\%  (50\%) fine-tuning the lower bound is 11 TeV (16 TeV) for the RS model and 3.2 TeV (5.3 TeV) for the model with modified background~\footnote{It has previously been noted that one can fine-tune the fermion bulk-masses in RS in order to achieve $R_b$ in agreement with experiment~\cite{Djouadi:2006rk}. Our analysis shows that in the minimal anarchic RS model such a fine-tuning is sizable.}.

Finally we should mention that, to be fully consistent, one should consider a global fit of the EWPT data to the observables $S, T$ and $\delta\tilde g_{b_L}$ and also include possible loop corrections~\cite{Carmona:2011ib}. We will leave this to future work.

\section{Flavor violation}

\label{quarkflavorgeneral}

The dominant flavor violation comes from the KK gluons, in particular the off-diagonal elements in Eq.~(\ref{gKK}). Following standard convention~\cite{Bona:2007vi,Isidori:2010kg} we parametrize the most constraining $\Delta F=2$ Lagrangian as~\footnote{The minus signs in the first two operators reflect our convention for the metric, $\eta^{\mu\nu}=\diag(-+++)$.}
\bea
- \mathcal L_{sd}^{\Delta F=2}=
\mathcal H_{sd}^{\Delta F=2}&=&
-C_1^{sd} (\bar d_L\gamma^\mu s_L)^2-\tilde C_1^{sd}(\bar d_R\gamma^\mu s_R)^2\nn\\
  &&+\, C_4^{sd}(\bar d_L s_R)(\bar d_R s_L)
  +C_5^{sd}(\bar d_L^\alpha s_R^\beta)(\bar d_R^\beta s_L^\alpha)\,.
  \label{Wilson}
\eea
In full analogy we can write similar operators by replacing $sd\to uc$ or $bd$.
We can use the results of Sec.~\ref{KKgluons} to write the coefficients explicitly as
\bea
C_1^{sd}&=&\frac{g_s^2\, y_1}{6}\int e^{2A}(\Omega_{d_L}^{12})^2\label{C1}\,,\\
\tilde C_1^{sd}&=&\frac{g_s^2\, y_1}{6}\int e^{2A}(\Omega_{d_R}^{12})^2\label{C1tilde}\,,\\
C_4^{sd}&=&-g_s^2\, y_1\int e^{2A}(\Omega_{d_L}^{12}\Omega_{d_R}^{12})
\label{C4}\,,\\
C_5^{sd}&=&\frac{g_s^2\, y_1}{3}\int e^{2A}(\Omega_{d_L}^{12}\Omega_{d_R}^{12})\,,\label{C5}
\eea
where $\Omega^{12}_{d_\chi}$ has been defined in Eq.~(\ref{Omega}).
Notice that using the unitarity of the mixing matrices we can write
\be
\Omega_{d_L}^{12}=
  (\Omega_{d_L}^2-\Omega_{d_L}^1)V_{d_L}^{12}V_{d_L}^{*22}
 +(\Omega_{d_L}^3-\Omega_{d_L}^1)V_{d_L}^{13}V_{d_L}^{*23}\,,
\ee
and similarly for $L\to R$. 

\begin{figure}[htb]
\begin{center}
\begin{psfrags}
\input{C4PDF-psfrag.tex}
\includegraphics[width=0.48\textwidth]{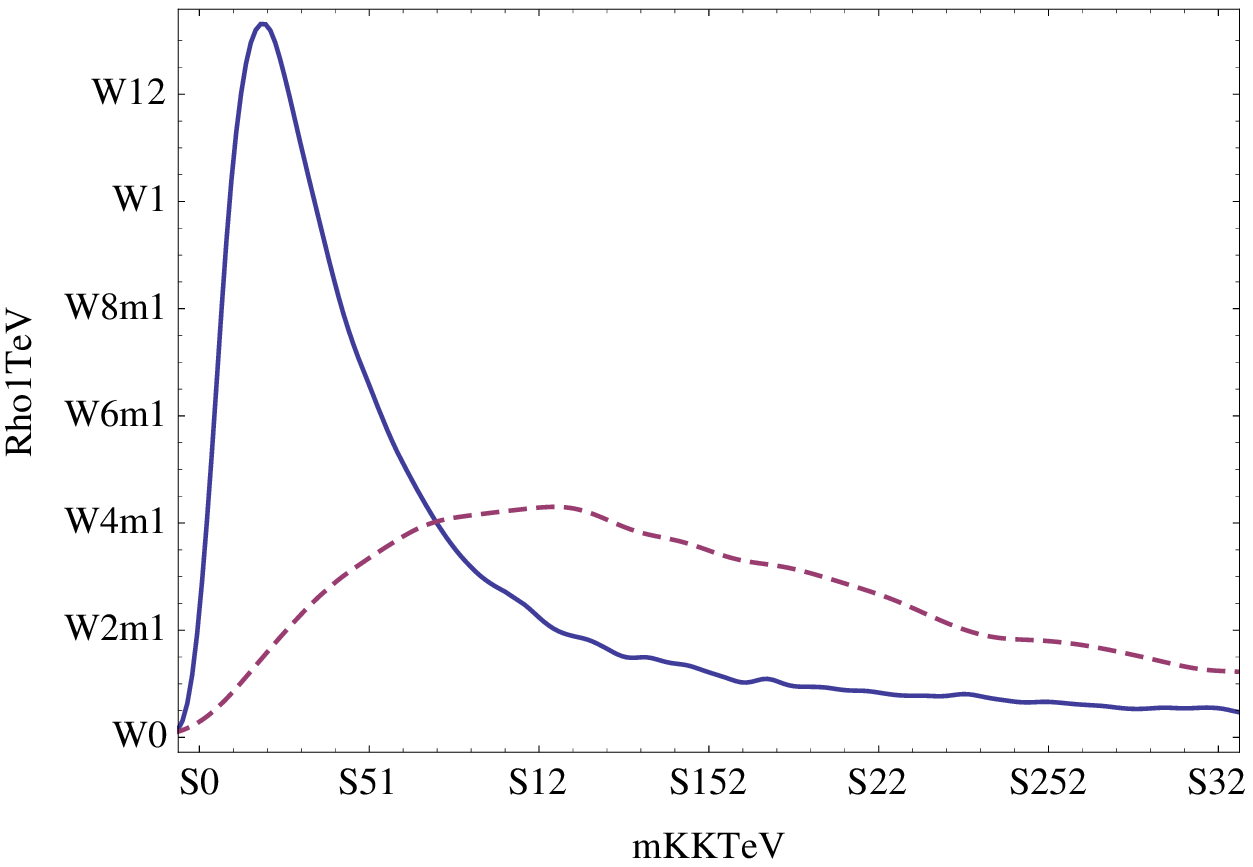}
\end{psfrags}
~
\begin{psfrags}
\input{C4CDF-psfrag.tex}
\includegraphics[width=0.48\textwidth]{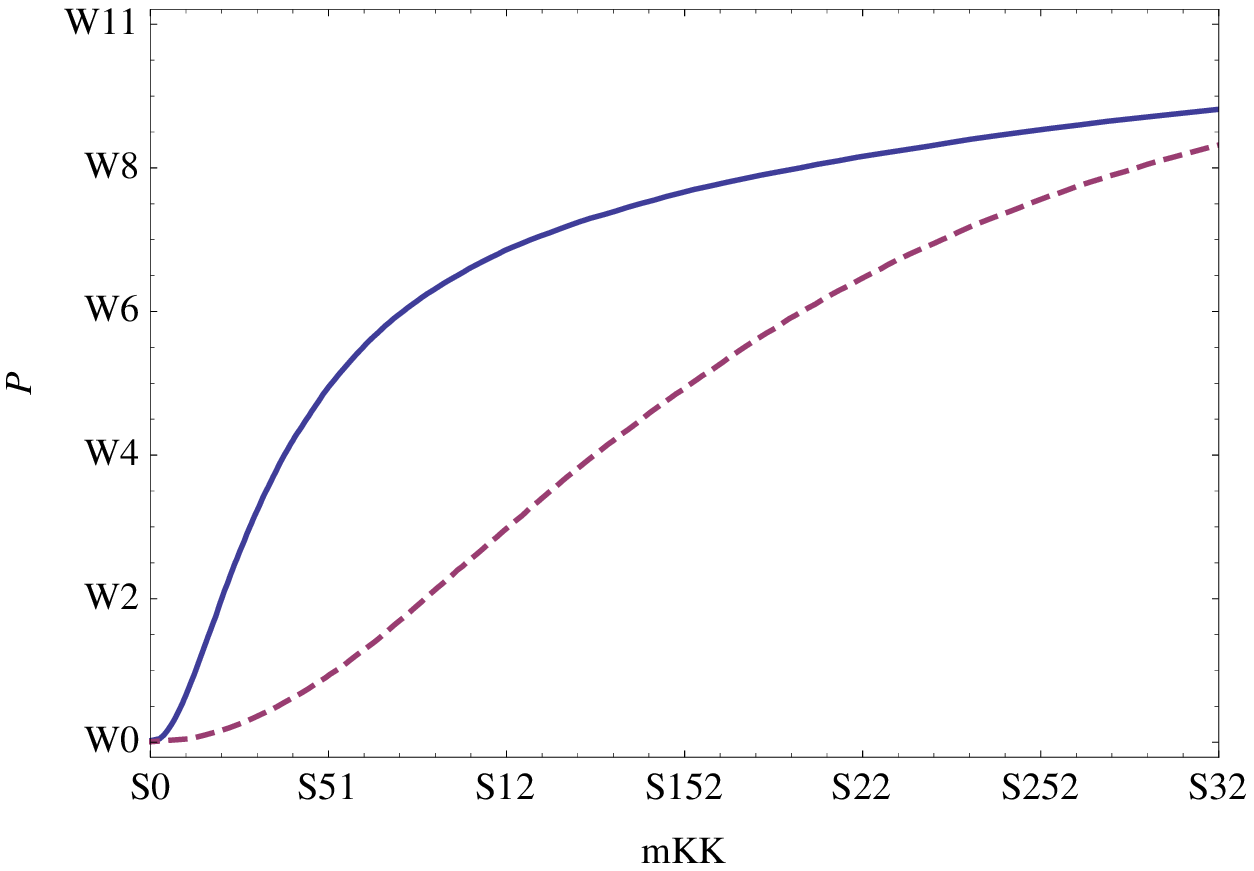}
\end{psfrags}
\end{center}
\caption{\it PDF (left panel) and CDF (right panel) for $m_{KK}$ from comparison with $\left|\textrm{Im} \,C_4^{sd}\right|$. Dashed lines correspond to the RS model and solid lines to the model in Eq.~(\ref{ourmetric}) for $k\Delta=1$ and $\nu=0.5$.}
\label{figC4}
\end{figure}

The coefficients $C_i$ are related to flavor violating and/or $CP$ violating observables~\cite{Isidori:2010kg} from where they get upper bounds. With our set of data points we can then compute the exact mixing matrices numerically and use them to find the coefficients $C_i$ defined above. The former bounds are then translated into lower bounds on $m_{KK}$ (the mass of the first resonance of the gluon). The most constraining parameter is Im $C_4^{sd}$, which is related to the $CP$ violating observable in the $K$-system, $\epsilon_K$, and is bounded by~\cite{Isidori:2010kg}
\be
\left|\textrm{Im}\, C_4^{sd}\right|<2.6\times 10^{-11}\ \textrm{TeV}^{-2}\,.
\label{ImC4}
\ee
By using the expression for $C_4^{sd}$ provided in Eq.~(\ref{C4}) and comparing with the experimental bound in Eq.~(\ref{ImC4}) we obtain bounds on $m_{KK}$ for every data point~\footnote{A more refined procedure would be to link the Wilson coefficients in Eq.~(\ref{Wilson}) to the actual observables, in particular $\epsilon_K$ and $\Delta m_K$, and apply the direct experimental bounds, see e.g.~Ref.~\cite{Huber2}.}, as we did in the previous section for the coupling $R_b$. The result is exhibited in Fig.~\ref{figC4} where we show both the PDF and CDF for the distribution of points.

A statistical analysis similar to that done in Sec.~\ref{EWPT} can be performed here,
and in Tab.~\ref{tabC4} 
\begin{table}[htb]
\centering
\begin{tabular}{|c|ccc|ccc|}
\cline{2-7}
\multicolumn{1}{c|}{}&\multicolumn{3}{c|}{\textit{Probability for $m_{KK}$ below}}&\multicolumn{3}{c|}{\textit{Percentile}}\\
\multicolumn{1}{c|}{}&$3$ TeV &$5$ TeV &$10$ TeV&10\% &20\% &50\% \\
\hline
RS 			& 2.4\% &6.4\% &22\% & 6.5 TeV & 9.7 TeV & 19 TeV\\
$\nu=0.5$ 	& 26\% &43\% &64\%   & 1.6 TeV & 2.5 TeV & 6.2 TeV\\
\hline
\end{tabular}
\caption{\it Left panel: Integrated probability for values of $m_{KK}$ below 3, 5 and 10 TeV from Im $C_4^{sd}$ for RS (upper row) and the model in Eq.~(\ref{ourmetric}) for $k\Delta=1$ and $\nu=0.5$ (lower row). Right panel: 10th, 20th and 50th percentiles for both models.}
\label{tabC4}
\end{table}
we present some explicit numbers obtained from these distributions. 
We can trace back the improvement in the bounds on $m_{KK}$ in the modified background model with respect to the RS model on the weakening of couplings of gauge KK modes to the first and second generation SM fermions, 
resulting in turn from the enhancement in the coefficients $c_\psi$. 
For instance assuming a 20\% (50\%) fine-tuning the lower bound for the RS model is 9.7 TeV (19 TeV) while for the modified background model they are 2.5 TeV (6.3 TeV). The combined bounds will be much stronger as we will show in the Sec.~\ref{conclusions}. 

Finally we should pay attention to the other coefficients: ${\rm Re\ }C^{sd}_4$, $C^{sd}_1$, $\tilde C^{sd}_1$ and $C^{sd}_5$. The bounds on ${\rm Re\ }C^{sd}_4$, coming mostly from $\Delta m_K$, are about one to two orders of magnitude weaker than for ${\rm Im\ }C^{sd}_4$. However it is conceivable that the favorable points that allow a low KK scale could result from an accidental cancellation of the phase and hence the bounds from the real part turn out to dominate. We have verified that this is not the case and the bounds are not changed by taking into account the real part. Furthermore, notice that $C^{sd}_5=\frac{1}{3}C^{sd}_4$ and hence whenever $C^{sd}_4$ is suppressed so is $C^{sd}_5$ (the experimental constraints on the two quantities are comparable). The experimental constraints on the coefficients $C^{sd}_1$ and $\tilde C^{sd}_1$ are about two orders of magnitude weaker with again a similar suppression as $C_4^{sd}$. We thus do not expect any additional constraints from here either.

\section{Summary and conclusions}
\label{conclusions}
In this paper we have considered a general 5D warped model, with SM fields propagating in the bulk of the fifth dimension, and computed explicit expressions for oblique ($S,\,T,\,W,\,Y$) and non-oblique  ($\delta \tilde g_{b_L}$) observables, as well as flavor and $CP$ violating effective four-fermion operators. We have worked out in particular the RS model and the model with the modified metric (\ref{ourmetric}). While there is a wide literature on the RS model, for the model of Eq.~(\ref{ourmetric}), introduced as an alternative to models with an extra gauge custodial symmetry in the bulk, only electroweak observables where cosidered \cite{Cabrer:2010si,Cabrer:2011vu,Carmona:2011ib} while its flavor structure was largely unexplored.  We have then concentrated here on bounds on $m_{KK}$ from the modification of the $Z\bar bb$ coupling and from FCNC and $CP$ violating operators. We have in all cases compared the result for the RS model with those for the model with metric (\ref{ourmetric}) and parameter values for which corrections to oblique observables are well under control for $m_{KK}\gtrsim 1$ TeV. 

We have randomly generated 40,000 sets of values for the 5D Yukawa couplings and made for each of them a $\chi^2$ fit to the quark mass eigenvalues and CKM elements for both models [RS and the model  (\ref{ourmetric}) with $k\Delta=1$ and $\nu=0.5$] generating in each case the fermion profiles such that we can then compare the results in both models.

Concerning non-oblique versus FCNC and $CP$ violating observables in both models the final comparison is as follows:
\begin{itemize}
\item
The bounds for the modified metric model are milder than those in the RS model. This can be clearly seen from Figs.~\ref{figZbb} and \ref{figC4} and from Tabs.~\ref{tabZbb} and \ref{tabC4}. The main origin of this improvement in the modified metric model with respect to the RS model can be traced back to the fact that because of the IR deformation of the metric fermions fitting the quark mass eigenvalues and CKM matrix elements are shifted towards the UV in the former model which produced a general suppression of effects in the observables. 
\item
The bounds from FCNC and $CP$ violating effective operators are stronger than those from non-oblique observables in both models. This is mainly due to the strong constraints on these operators, in particular from the $CP$ violating observable $\epsilon_K$. Of course we expect  the combined bonds to be stronger than those from the individual constraints.
\end{itemize}
\begin{figure}[htb]
\begin{center}
\begin{psfrags}
\input{BothPDF-psfrag.tex}
\includegraphics[width=0.48\textwidth]{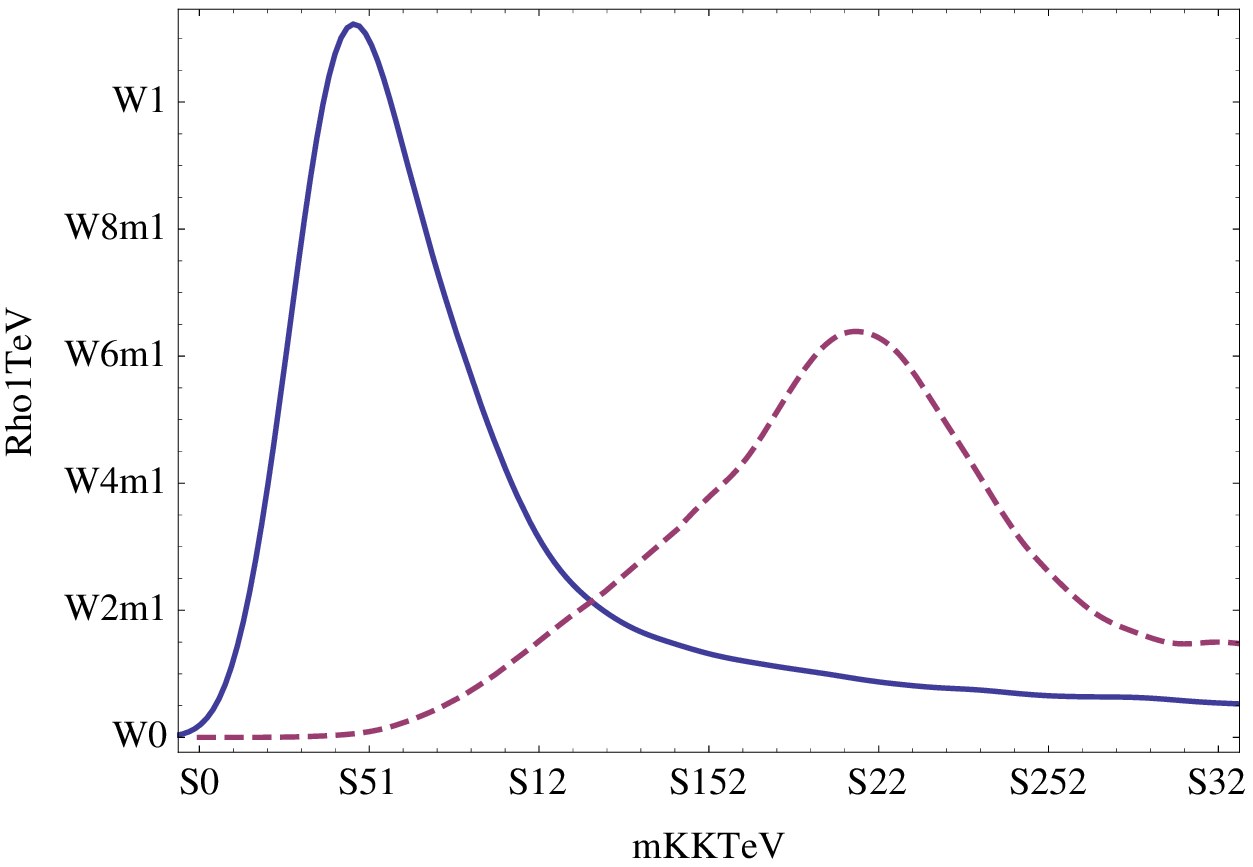}
\end{psfrags}
~
\begin{psfrags}
\input{BothCDF-psfrag.tex}
\includegraphics[width=0.48\textwidth]{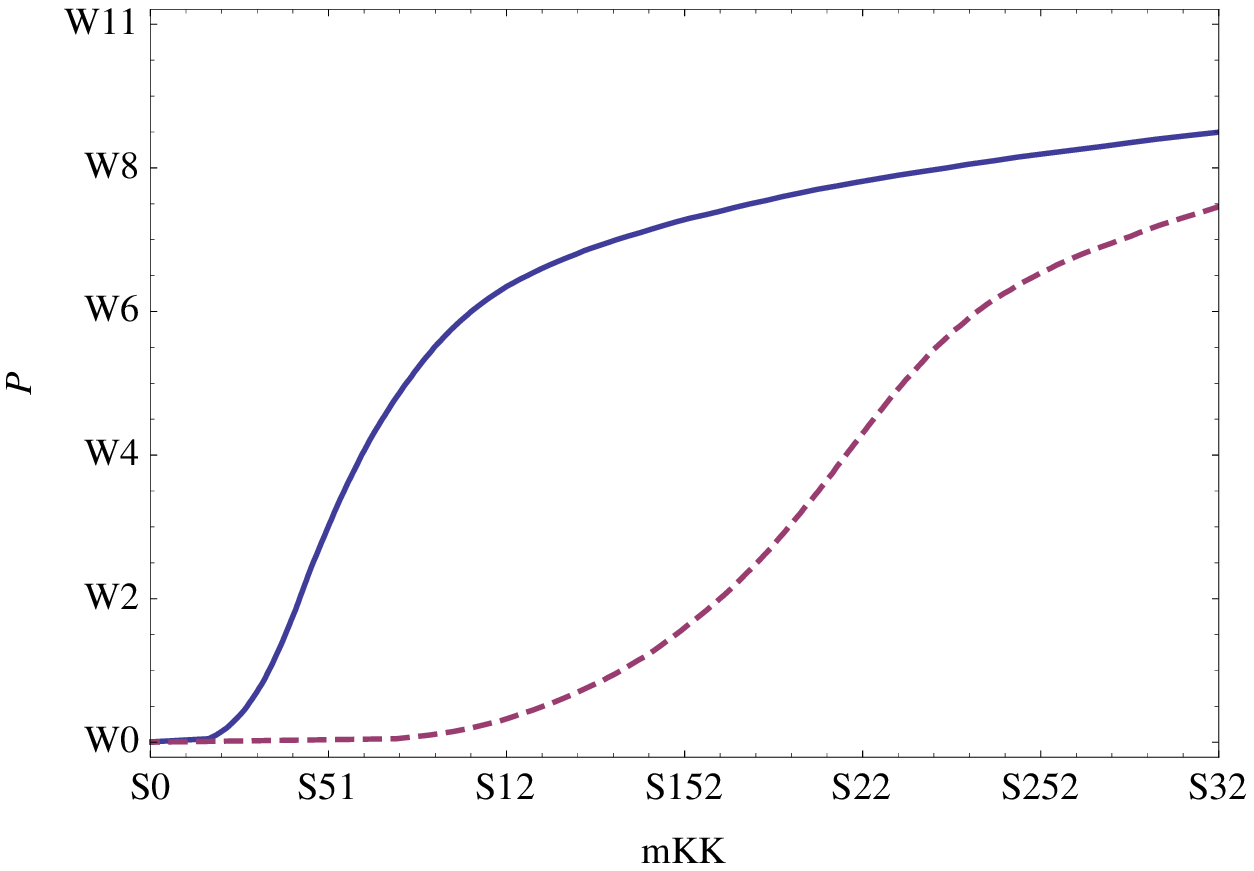}
\end{psfrags}
\end{center}
\caption{\it PDF (left panel) and CDF (right panel) for $m_{KK}$ from comparison with $\left|\textrm{Im} \,C_4^{sd}\right|$ and $R_b$. Dashed lines correspond to the RS model and solid lines to the model in Eq.~(\ref{ourmetric}) for $k\Delta=1$ and $\nu=0.5$.}
\label{figboth}
\end{figure}
In Fig.~\ref{figboth} we show the PDF and CDF distributions corresponding to the combined bounds from non-oblique observables and flavor/$CP$ violating effective operators. A similar statistical analysis to those presented for the individual contributions is done here and the results are presented in Tab.~\ref{tabboth}.
\begin{table}[htb]
\centering
\begin{tabular}{|c|ccc|ccc|}
\cline{2-7}
\multicolumn{1}{c|}{}&\multicolumn{3}{c|}{\textit{Probability for $m_{KK}$ below}}&\multicolumn{3}{c|}{\textit{Percentile}}\\
\multicolumn{1}{c|}{}&$3$ TeV &$5$ TeV &$10$ TeV&10\% &20\% &50\% \\
\hline
RS &0\% &0\% &3.3\% &13 TeV&16 TeV& 21 TeV\\
$\nu=0.5$ &7.1\% &30\% &64\% &3.3 TeV&4.2 TeV& 7.2 TeV\\
\hline
\end{tabular}
 \caption{\it Left panel: Integrated probability for values of $m_{KK}$ below 3, 5 and 10 TeV from $R_b$ and Im $C_4^{sd}$ for RS (upper row) and the model in Eq.~(\ref{ourmetric}) for $k\Delta=1$ and $\nu=0.5$ (lower row). Right panel: 10th, 20th and 50th percentiles for both models.}
\label{tabboth}
\end{table}
From there we can see that  assuming a 20\% (50\%) fine-tuning the lower bound for the RS model is 16 TeV (21 TeV) while for the modified background model they are 4.2 TeV (7.2 TeV). Then since the percentile is also a measure of the fine-tuning we can conclude that if we tolerate a fine tuning $\sim$10\%-20\% a KK-mass $\sim$ 3 TeV can be roughly acceptable.

Let us remark that the derived bounds can be considered the most conservative ones (i.e. the worst case scenario in the absence of further suppressions). In particular the $Z\bar bb$ bounds can be improved if one allows for a moderate hierarchy in the 5D Yukawas, i.e.~by lowering the 5D bottom Yukawa. On the other hand flavor bounds can be improved by including the effects of UV brane localized kinetic term for the gluon\cite{Csaki:2008zd}, or by invoking some flavor symmetries \cite{Santiago:2008vq}.

However before claiming a theory of quark flavor, a number of points, outside the scope of the present paper, should  be clarified. First of all the anarchic solution to the flavor problem should arise from an underlying UV completion, which should explain the values of the 5D Yukawa couplings and the localizing fermion coefficients $c_f$. In particular, all coefficients are quite close to the flat value $c_f=\frac{1}{2}$, owing to the fact that the quark mass hierarchies are actually much less than the Planck-weak hierarchy.
Second, all the results presented in this paper are tree-level results while radiative corrections should also be considered~\footnote{Radiative corrections to electroweak observables are already taken into account in Ref.~\cite{Carmona:2011ib}.}. They should depend to a large extent on the size of 5D Yukawa couplings so that they are very model dependent. Moreover there are some $CP$ violating effects which appear only at the loop level and which consequently we have not considered either here. In particular the one-loop contribution to the neutron electric dipole moment~\cite{Agashe:2004cp} due to non-removable Majorana phases would probably require some kind of flavor alignment although a bulk Higgs should certainly alleviate the problem since it renders the one-loop diagram contributing to it finite.
Finally a theory of lepton flavor can certainly be constructed along similar lines. We expect corresponding improvements for processes like $\mu\to e\gamma $ and $\mu\to 3e$, see also Ref.~\cite{Archer:2011bk}. 

\section*{Acknowledgments}

GG would like to thank S. Huber and S. J\"ager for discussions. Work supported in part by the Spanish Consolider-Ingenio 2010 Programme CPAN (CSD2007-00042) and by CICYT-FEDER-FPA2008-01430. The work of JAC is supported by the Spanish Ministry of Education through a FPU grant. The research of GG is supported by the ERC Advanced Grant 226371, the ITN programme PITN- GA-2009-237920 and the IFCPAR CEFIPRA programme 4104-2.

\newpage%
\vspace{2cm}
\appendix
\noindent
\textbf{\Large Appendix}
\section{Four-fermion terms from the EW KK modes}
\label{4f}
In this appendix we explicitely write the four-fermion interactions. For the neutral currents the effective Lagrangian reads
\be
\mathcal L^{4f}_{NC}=\sum_{f_\chi,f'_\chi}
\delta^{k\ell,rs}_{f_\chi,f'_\chi}
(\bar f^k_\chi\gamma^\mu f^\ell_\chi) (\bar f'^r_{\chi'}\gamma_\mu f'^s_{\chi'})
\label{4fNC}\,,
\ee
where the constants $\delta$ are tensors in flavor space:
\be
\delta^{k\ell,rs}_{f_\chi,f'_\chi}=
\frac{e^2}{2}\left(Q_fQ_{f'}+\frac{1}{s_W^2c_W^2}g^{SM}_{f_\chi}g^{SM}_{f'_\chi}\right)
\sum_{i,j}\hat\alpha_{f^i_\chi,f'^{j}_\chi}(V^{k i}_{f_\chi}
V^{*\ell i}_{f_\chi})(V^{rj}_{f'_\chi}V^{*sj}_{f'_\chi})\,.
\ee
Finally, integrating out the KK modes of the $W$ boson also leads to four-fermion terms, which we write explicitly as
\bea
\mathcal L^{4 f}_{CC}&=& \delta_{u^kd^\ell,d^r u^s} (\bar u^k_L\gamma^\mu d_L^\ell)  (\bar d_L^r\gamma^\mu u_L^s)
+\delta_{\nu^ke^\ell,e^r \nu^s} (\bar \nu^k_L\gamma^\mu e_L^\ell)  (\bar e_L^r\gamma^\mu \nu_L^s)\nn\\
&&+\left[\delta_{u^kd^\ell,e^r \nu^s} (\bar u^k_L\gamma^\mu d_L^\ell)  (\bar e_L^r\gamma^\mu \nu_L^s)+h.c.\right]\,,
 \label{4fCC}
\eea
with
\bea
\delta_{u^kd^\ell,d^r u^s}&=&\frac{g^2}{2}\sum_{ij}\hat\alpha_{q_L^i,q^j_L}
   V_{u_L}^{ki}V_{d_L}^{*\ell i}V^{rj}_{d_L}V^{*sj}_{u_L}\,,\nn\\
\delta_{\nu^ke^\ell,e^r \nu^s}&=&\frac{g^2}{2}\sum_{ij}\hat\alpha_{\ell_L^i,\ell^j_L}
   V_{\nu_L}^{ki}V_{e_L}^{*\ell i}V^{rj}_{e_L}V^{*sj}_{\nu_L}\,,\nn\\
\delta_{u^kd^\ell,e^r \nu^s}&=&\frac{g^2}{2}\sum_{ij}\hat\alpha_{q_L^i,\ell^j_L}
   V_{u_L}^{ki}V_{d_L}^{*\ell i}V^{rj}_{e_L}V^{*sj}_{\nu_L}\,.
\eea

\section{Fermion propagator}
\label{app:fermions}

In this appendix  we will provide a few more details concerning the procedure of integrating out fermionic KK modes, as done in Sec.~\ref{secKKfermions}. This parallels and generalizes the computation in Ref.~\cite{Cabrer:2010si} for the gauge bosons. We will restrict ourselves to the case where the KK tower contains a zero mode that has to be subtracted, which is the most complicated case and the only relevant for this work. In fact, in order to evaluate the first diagram in Fig.~\ref{KKfermion} we need to compute
\be
\beta^{d_L}_{i\ell}=\sum_j\hat Y^d_{ij}\hat Y^{d*}_{\ell j}\sum_{n\neq 0}\int_0^{y_1}dy\,dy'
\,\frac{
 \left[\xi^0(y)\,\hat \psi^0_{Q_L^i}\!(y)\,\hat\psi^n_{d_R^j}\!(y)\right]
 \left[\xi^0(y')\,\hat \psi^0_{Q_L^\ell}\!(y')\,\hat\psi^n_{d_R^j}\!(y')\right]}{m_n^2}\,,
\label{betaexpl}
\ee
where $\xi^0(y)$ is the normalized Higgs zero mode wave function. The expression for $\beta^{d_R}_{i\ell}$ is completely analogous. The idea is now to first perform the sum over the KK modes and then the integrations.
Let us thus consider the equations of motion for the KK modes of a fermion,
\be
e^{2A} m_n^2 \hat \psi_n+e^A(\partial_y-Q')e^{-A}(\partial_y+Q')\hat \psi_n=0\,,
\qquad 
\left.\hat \psi_n'+Q'\hat\psi_n\right|_{y=0,y_1}=0\,,
\label{eomfermion1}
\ee
which follow from Eq.~(\ref{Dirac2}).
It will be convenient to factor out the zero mode as
\be
\hat\psi_n(y)=e^{-Q(y)}\chi_n(y)
\ee
which transforms Eq.~(\ref{eomfermion1}) into
\be
e^{A-2Q}\,m_n^2\,\chi_n+(e^{-A-2Q}\chi'_n)'=0\,,\qquad \left.\chi_n'\right|_{y=0,y_1}=0\,.
\label{eomfermion}
\ee
The completeness and orthonormality conditions in this basis read:
\be
\sum_{n=0}^{\infty} \chi_n(y)\chi_n(y')=e^{-A+2Q}\delta(y-y')\,,\qquad
\int_0^{y_1}e^{A(y)-2Q(y)}\chi_n(y)\chi_m(y)=\delta_{mn}\,. \label{completeness}
\ee
We now need to compute the sum
\be
G(y,y')=\sum_{n\neq 0}\frac{\chi_n(y)\chi_n(y')}{m_n^2}\,.
\ee 
Note that the zero mode has been excluded from the sum.
To compute $G(y,y')$, we integrate Eq.~(\ref{eomfermion}) twice
\be
\chi_n(y)=\chi_n(0)-m_n^2\int^y_0 e^{A(u)+2Q(u)} \int_0^{u}  e^{A(v)-2Q(v)}\chi_n(v)\,,
\ee
and use Eq.~(\ref{completeness}) to get
\be
G(y,y')
=\sum_{n\neq0}\frac{\chi_n(0)\chi_n(0)}{m_n^2}
-\int_0^{y_>}e^{A+2Q}(1-\Omega)
+\int_0^{y_<}e^{A+2Q}\Omega\,,
\ee
where $\Omega(y)$ has been defined in Eq.~(\ref{Omegafermion}), and $y_>$ ($y_<$) is the larger (smaller) of the pair $(y,y')$. We thus have reduced the problem of finding $G(y,y')$ to that of finding $G(0,0)$, which is the zero momentum limit of the (zero mode subtracted) brane-to-brane propagator. The latter can be written as 
\be
G(0,0;p^2)=-\frac{\chi(0,p^2)}{\chi'(0,p^2)}-\frac{\chi^2_0(0)}{p^2}\,
\ee
where $\chi(0,p^2)$ is the solution to
\be
-e^{A-2Q}p^2\chi+(e^{-A-2Q}\chi')'=0\,,\qquad \left.\chi'\right|_{y_1}=0\,.
\label{eomfermion2}
\ee
(note that we do not impose a BC at $y=0$). One can easily derive an equation for $\chi'/\chi $ and solve it in a power series in $p^2$:
\be
e^{-A(y)-2Q(y)}
\frac{\chi'(y,p^2)}{\chi(y,p^2)}=-p^2\int_y^{y_1}e^{A(u)-2Q(u)}+p^4\int_y^{y_1}e^{A(u)+2Q(u)}\left[\int_u^{y_1}e^{A(v)-2Q(v)}\right]^2+\dots
\ee
One ends up with
\be
G(0,0)=\int_0^{y_1}e^{A+2Q}(1-\Omega)^2\,,
\ee
and hence
\be
G(y,y')
=\int_0^{y_1}e^{A+2Q}(1-\Omega)^2
-\int_0^{y_>}e^{A+2Q}(1-\Omega)
+\int_0^{y_<}e^{A+2Q}\Omega\,.
\ee
Using this expression in Eq.~(\ref{betaexpl}) we arrive, after a series of partial integrations, at the quoted result Eq.~(\ref{beta}).

\section{Right handed hierarchies}
\label{RH}

In Sec.~\ref{quarks} we have given expressions for the masses and left handed mixing angles in case there is a left handed hierarchy, $Y^q_{1i}\ll Y^q_{2i}\ll Y^q_{3i}$. This fact is well supported by experiment, given that the CKM mixing angles are hierarchical. There is no such analogous measurement for the right handed mixing angles. However, making the assumptions that we also have a right-handed hierarchy,
\be
Y^q_{i1}\ll Y^q_{i2}\ll Y^q_{i3}\,,
\label{RHhierarchy}
\ee
the expressions given in Sec.~\ref{quarks} simplify. Although the calculation is a bit tedious, the result is very simple: we just have to replace the mass-squared matrices by the Yukawas. Indeed,
by writing the expressions in Eq.~(\ref{angulos}) and Eq.~(\ref{masascomp}) explicitly in terms of the 
Yukawa couplings and taking the limit Eq.~(\ref{RHhierarchy}) we obtain for the angles
\begin{align}
V^{q_L}_{12}&=-\widetilde Y^q_{12}\,/\,\widetilde Y^q_{22}\,,\qquad\qquad
  V^{q_L}_{21}=(\widetilde Y^{q}_{12}\,/\,\widetilde Y^q_{22})^*\,,\nn\\
V^{q_L}_{23}&=-Y^q_{23}\,/\,Y^q_{33}\,,\qquad\qquad
  V^{q_L}_{32}=(Y^q_{23}\,/\,Y^q_{22})^*\,,\nn\\
V^{q_L}_{13}&=-Y^q_{13}\,/\,Y^q_{33}
+(\widetilde Y^q_{12}Y^q_{23})/(\widetilde Y^q_{22}Y^q_{33})\,,\nn\\
V^{q_L}_{31}&=(Y^q_{13}\,/\,Y^q_{33})^*\,,
\label{angulos2}
\end{align}
and for the mass eigenvalues
\bea
(m_3^q)^2&=&\frac{v^2}{2}\,|Y^q_{33}|^2,\nonumber\\
(m_2^q)^2&=&\frac{v^2}{2}\,|\widetilde Y_{22}^{q}|^2\,,\nonumber\\
(m_1^q)^2&=&\frac{v^2}{2}\,|Y_{11}^{q}-\widetilde Y_{12}^{q}\widetilde Y_{21}^{q}/\widetilde Y^{q}_{22}|^2\,,
\label{masascomp2}
\eea
where we have defined
\be
\widetilde Y^{q}_{ij}=Y^{q}_{ij}-\frac{Y^{q}_{i3}Y^{q}_{3j}}{Y_{33}}\,.
\ee
These results agree with the ones quoted in Ref.~\cite{Hall} whose authors considered real Yukawas. In the case of a right handed hierarchy, there is also an approximation to the right handed rotations. It can be obtained from Eq.~(\ref{angulos2}) by replacing $Y^q\to Y^{q\dagger}$, leading to expressions again in agreement with those in Ref.~\cite{Hall}.

\section{EWPO effects on couplings}
\label{EWPO}

Following Ref.~\cite{Burgess:1993vc} we proceed as follows
\begin{enumerate}
\item
Diagonalize and canonically normalize the kinetic terms for the gauge bosons. To simplify a little, we will use the fact that $W=Y$, which is equivalent to not having any mixing between Z and $\gamma$ at $\mathcal O(p^4)$.
\item
Express the SM input parameters $\tilde e$, $\tilde s_w$ and $\tilde m_Z$ appering in $\mathcal L_{SM}$ in terms of the physically measured ones $e$, $s_w$ and $m_Z$. The latter are inferred from the measurements of $\alpha$, $G_F$ and $m_Z$. Beyond the contributions identified in Ref.~\cite{Burgess:1993vc}, the only difference is a shift in the $Z$ mass due to the presence of $Y$ and $W$, while $\alpha$ and $G_F$ remain unchanged~\footnote{We use the fact that we are in the oblique basis, where the charged current vertex corrections of electrons and muons as well as the four fermi muon-decay operator are zero.}.
\end{enumerate}
We then find the following corrections to the SM gauge couplings~\footnote{
We recall the SM values  $g^{SM}_{f_L}=T^3_f-s_w^2 Q^{em}_f $, $g^{SM}_{f_R}=-s_w^2 Q^{em}_{f}$ with $T^3_f=\{\frac{1}{2},-\frac{1}{2},-\frac{1}{2},\frac{1}{2}\}$ and $Q^{em}_f=\{\frac{2}{3},-\frac{1}{3},-1,0\}$ for $f=\{u,d,e,\nu\}$.}:
\bea
\delta g_{f_\chi}&=&\frac{g^{SM}_{f_\chi}}{2} \left(\alpha T+\frac{Y}{c_w^2}\right)
-Q^{em}_f\,\frac{1}{c_w^2-s_w^2}\left(\frac{\alpha S}{4}-c_w^2 s_w^2\,\alpha T-s_w^2\,Y\right)+\delta\tilde g_{f_\chi}\,,\nn\\
\label{deltag}\\
\delta  h_{ud}&=&-V_{CKM}\,\frac{1}{2(c_w^2-s_w^2)}
\left(\frac{\alpha S}{2}-c_w^2\,\alpha T-Y\right)+\delta \tilde h_{ud}\,,\\
\delta  h_{\nu e}&=&-V_{PMNS}\,\frac{1}{2(c_w^2-s_w^2)}
\left(\frac{\alpha S}{2}-c_w^2\,\alpha T-Y\right)+\delta \tilde h_{\nu e}\,,
\eea
where the tilded quantities refer to the explicit vertex corrections stemming from the various non-oblique corrections (see below for the explicit expressions).
The tree level inverse propagators  for the gauge bosons now take the simple diagonal form
\bea
\Pi_\gamma(s)&=&s\left[1+\frac{Y}{c_w^2}\,\frac{s}{m_Z^2}\right]\,,\\
\Pi_Z(s)&=&(s-m_Z^2)\left[1+\frac{Y}{c_w^2}\left(1+\frac{s}{m_Z^2}\right)\right]\,,\\
\Pi_W(s)&=&(s-m_W^2)\left[1+Y\left(1+\frac{s}{m_W^2}\right)\right]\,.
\eea
In the last propagator the physical $W$ mass can be expressed as
\be
m_W^2=m_Z^2c_w^2\left(1-\frac{\alpha S}{2(c_w^2-s_w^2)}+\frac{c_w^2\,\alpha T}{c_w^2-s_w^2}+\frac{s_w^2\,Y}{c_w^2-s_w^2}\right)\,.
\ee
The four-fermion terms in Eq.~(\ref{4fNC}) and Eq.~(\ref{4fCC}) remain unchanged by this procedure.

\end{document}